\documentclass[aps,prd,twocolumn,amsmath,superscriptaddress,amssymb,showpacs,floatfix,nofootinbib,longbibliography]{revtex4-1}

\usepackage{graphicx}
\usepackage{dcolumn}
\usepackage{xcolor}
\usepackage{bm}

\usepackage{calligra}
\usepackage[T1]{fontenc}
\usepackage{egothic}
\usepackage[T1]{fontenc}
\newfont{\rsfsten}{rsfs10 scaled 1200}
\newfont{\rsfsseven}{rsfs10 scaled 1200}
\newfont{\rsfsfive}{rsfs10 scaled 1200}
\usepackage{epsfig}
\usepackage{units}
\usepackage[utf8]{inputenc}
\usepackage{hyperref}
\usepackage{tabularx}

\newcommand{\be}{\begin{equation}}
\newcommand{\ee}{\end{equation}}
\newcommand{\bea}{\begin{eqnarray}}
\newcommand{\eea}{\end{eqnarray}}

\usepackage{comment}

\begin{document}

\title{On the mass distribution of the LIGO-Virgo-KAGRA events}

\author{Mehdi El Bouhaddouti}
\email{melbouhaddouti@oakland.edu, ORCID: orcid.org/0009-0001-1299-0879}
\affiliation{Department of physics, Oakland University, Rochester Michigan, 48309,U.S.A}

\author{Ilias Cholis}
\email{cholis@oakland.edu, ORCID: orcid.org/0000-0002-3805-6478}
\affiliation{Department of physics, Oakland University, Rochester Michigan, 48309,U.S.A}

\begin{abstract}
The merging black hole binaries detected by the LIGO-Virgo-KAGRA (LVK) gravitational-wave observatories, may help us shed light on how such binaries form. In addition, these detections can help us probe the hypothesized primordial black holes, a candidate for the observed abundance of dark matter. In this work, we study the black-hole mass distribution obtained from the LVK binary black hole merger events. 
In particular, we study the primary mass $m_{1}$-distribution of the observed black hole binaries, and also the secondary to primary mass ratio $q = m_{2}/m_{1}$ distribution. 
We obtain those distributions by first associating a skewed normal distribution to each event detected with a signal to noise ratio (SNR) $>$ 8 and then summing all such distributions. We also sample black hole binaries from two separate populations of merging binaries to which we associate a redshift from the redshift distribution. One of these is a stellar-origin population that follows a mass-distribution similar to the zero-age mass function of stars. The second population of black holes follows a Gaussian mass-distribution. Such a distribution could approximate a population of black hole binaries formed from earlier black hole mergers in dense stellar environments, or binaries of primordial black holes among other astrophysical processes. For those populations, we evaluate the number of detectable events and fit their combination to the LVK observations. 
In our work, we assume that stellar-origin binary black holes follow a similar mass-distribution to that of the initial mass-function of stars. We simulate a wide range of stellar-origin
black-hole mass distributions.
In agreement with the binary black hole merger rates-analysis of the LVK collaboration, we find that studying the the observed LVK events can be fitted better by the combination of such a stellar-origin mass distribution and a Gaussian distribution, than by the stellar-origin mass distribution alone. As we demonstrate with some simple examples, our methodology allows for rapid testing of potential theoretical models for the binary black hole mergers to the observed events by LVK. 

\end{abstract}

\maketitle

\section{Introduction} \label{sec:intro}

The Laser Interferometer Gravitational-wave Observatory (LIGO), first detected gravitational waves (GW) from the coalescence of binary compact objects in 2015 \cite{LIGO_first_event}.
Since then, following detector upgrades and the Virgo \cite{Accadia_2012} and Kamioka Gravitational Wave Detector (KAGRA) \cite{10.1093/ptep/ptaa125} sites in Italy and Japan joining in the observations (Virgo in 2017 and KAGRA in 2019), there have been two more completed observing runs, O2 from November 2016 to August 2017 \cite{PhysRevX.9.031040} and O3 from April 2019 to the end of March 2020\cite{PhysRevX.13.011048}. As a result, the catalog of binary compact object mergers detected by the LIGO-Virgo-KAGRA collaboration (LVK) has reached 93~\cite{Abbott_2023}, with the O4 run currently ongoing. 
For each transient signal, the LVK collaboration evaluates the signal to noise ratio (SNR) and selects as merger events those with SNR>8.  
For every merging binary, LVK estimates the mass of the more massive compact object $M_1$ and of the less massive object $M_2$, their ratio $q = \frac{M_2}{M_1}$ and the redshift $z$ of the event, among other quantities.
 This allows for a study of the mass distribution of merging black holes (BHs). Among the 93 events in the LVK collaboration data, there are binary black holes (BBHs), binary neutron stars and neutron star-black hole binaries. In this paper we focus on the BBHs.

The two LIGO detectors at Hanford Washington and Livingston Louisiana, having the better sensitivity, offer the main means by which GWs from merging binaries are detected. 
Virgo and KAGRA allow for an improved localization of the events. The LIGO detectors are versions of a Michelson interferometer with 4 km long arms \cite{2015}, capable of detecting GWs in frequencies of $10$ Hz to $1000$ Hz. Thus they are ideal for the detection of coalescing stellar-mass BBHs.   

BBHs can have two origins, astrophysical, from collapsing stars and primordial, forming in the early universe (see e.g. \cite{H_tsi_2021, Karathanasis:2022rtr, ulrich2024multigenerationalblackholepopulation}). 
Astrophysical first generation BHs can be expected to have an approximately power-law mass distribution, similar to that of their progenitor stars \footnote{Second generation BHs, coming from the merger of first generation astrophysical BHs (originating form the collapse of stellar cores), may have a more complex mass-distribution.}. For stars with a mass larger than $1M_\odot$ (including massive enough to form BHs after their collapse), their initial mass follows a power-law distribution of $dN/dM \sim M^{-2.3 \pm 0.7}$ \cite{Kroupa_2001}.  
In this paper, we use a power-law model to simulate astrophysical BHs and test the resulting distribution from our simulations with the data from the LVK collaboration,  by searching for the best fit power-law parameters (For detailed reviews on rates and formation channels of BBHs see \cite{Mandel_2022}\cite{Spera_2022} ).

Primordial black holes (PBHs), differ from astrophysical ones in their origin. PBHs would form from curvature fluctuations in the primordial universe that collapse to black holes. Those fluctuations/overdensities are model dependent. They can be related to a phase transition during inflation (see for example~\cite{arbey2024primordialblackholessmall}). Due to the 'no-hair' conjecture, BHs are defined by three quantities: mass, spin and charge. 
PBHs would be indistinguishable from astrophysical BHs. However, the dynamics of PBHs and astrophysical BHs are different, with astrophysical BH binaries closely 
following the baryons in galaxies while PBHs being in dark matter halos. 
Given their different origins and environments, astrophysical and primordial BHs constitute two separate populations. Testing PBHs is motivated given that if present, then they would compose at least a fraction of dark matter (DM) in the mass range where the current GW detectors are sensitive. Moreover, measuring their mass would shed light to the primordial curvature fluctuations in the early universe and indirectly probe an energy scale impossible to reach with high-energy particle colliders. 
In this work, we consider for simplicity that PBHs have a monochromatic mass-distribution. Given that the LVK detectors have a finite mass resolution, such a monochromatic mass distribution would appear in the data as a Gaussian-like component.

In this paper, we present a methodology in studying the LVK BH mass-distribution data. 
In Section~\ref{sec:method}, we discuss how we take the reported $M_{1}$ and $M_{2}$ values to construct $M_{1}$- and $q$-distributions for the BBHs observed with a false alarm rate (FAR) smaller than 1 yr$^{-1}$. This is approximately similar to a SNR$>8$. 
Subsequently, in Section~\ref{sec:Simulated_BBHs}, we simulate BBHs from the two populations and compare to the data. Our results are discussed in Section~\ref{sec:results}. We find that the LVK BBHs require the presence of a Gaussian-like component in addition to a power-law mass-distribution. 
Finally, in Section~\ref{sec:conc}, we present our conclusions and discuss how such an analysis can be used to probe for the origin of that Gaussian-like component. \\

\section{The distribution of binary black hole masses} \label{sec:method}
\subsection{Studied population}\label{subsec:pop_stud}
We use the observations from the LVK O1, O2 and O3 runs \cite{PhysRevX.13.011048} obtained from the gravitational wave open science center \cite{Abbott_2023}. Out of the 93 events in the catalog we keep the events that are mergers of BBHs. We use the primary mass $M_1$ and secondary mass $M_2 > 4 \, M_{\odot}$ information, for events with a FAR smaller than 1 $\textrm{yr}^{-1}$. These conditions restrict the number of events that we work on, to 72.

\subsection{Probability density function of primary mass and binary mass-ratio}

For our study, we need to obtain, for each of the 72 events, the probability density functions (PDFs) for the primary mass $M_1$ and the mass-ratio $q$ \cite{PhysRevX.13.011048}. The reported error-bars from LVK on each of these quantities are asymmetric. Thus, we choose a skew-normal distribution to fit the $M_{1}$ and $M_{2}$ PDFs for each event. We use the following definition of the skew-normal distribution \cite{Azzalini},
\begin{equation}
    f(x,\xi,\omega,\alpha) = \frac{1}{\sqrt{2\pi}\omega}exp(-\frac{(x-\xi)^2}{2\omega^2})(1 + erf(\frac{\alpha(x-\xi)}{\omega\sqrt{2}}).
\end{equation}
The parameter $\xi$ is used to describe the location along the $x$-axis, $\omega$ is the scale parameter and $\alpha$ is the shape parameter (for $\alpha=0$ we get the normal distribution).
We want to fit the PDF of an event's $M_1$ so that the mode, $Y(\xi,\omega,\alpha)$ of the distribution is located at the central value of $M_1$ from the LVK data. We also want the scale of the distribution to be consistent with the claimed uncertainty on $M_1$. There is no analytic expression for the mode of the distribution for $\alpha \neq 0$. However, we can obtain the approximate expression using \cite{Azzalini},
\begin{equation}
    \delta(\alpha) = \frac{\alpha}{\sqrt{1+\alpha^2}} 
\end{equation}
and 
\begin{equation}
    \gamma_1=\frac{4-\pi}{2}\frac{(\delta(\alpha)\sqrt{2/\pi})^3}{(1-\delta(\alpha)^2/\pi)^{3/2}}.
\end{equation}
$\gamma_1$ is the skewness parameter of the distribution. Also, setting parameters,
\begin{equation}
   \mu_z(\alpha)=\sqrt{\frac{2}{\pi}}\delta(\alpha), 
\end{equation}
\begin{equation}
    \sigma_z(\alpha)=\sqrt(1-\mu_z^2(\alpha))
\end{equation}
and
\begin{equation}
    m_0(\alpha)=\mu_z(\alpha)-\frac{\gamma_1\sigma_z(\alpha)}{2}-\frac{sign(\alpha)}{2}exp(\frac{-2\pi}{\alpha}),
\end{equation}
we get the mode of the distribution expressed as,
\begin{equation}
    Y(\xi,\omega,\alpha)=\xi+\omega m_0(\alpha).
\end{equation}

\begin{figure}[ht!]
    \begin{centering}
    \vspace{-0.25in}
    \includegraphics[width=0.52\textwidth]{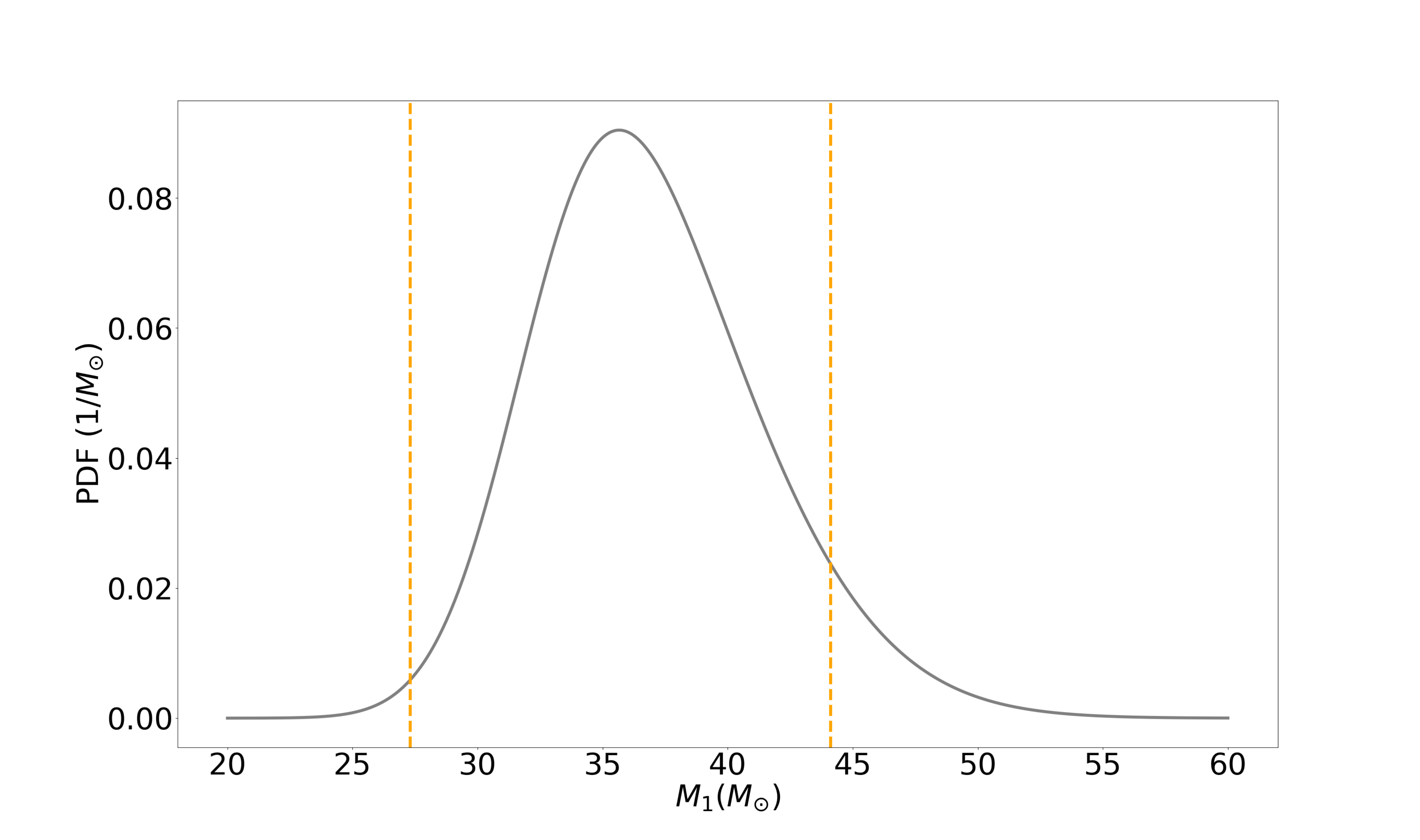}
    \end{centering}
    \vspace{-0.2in}
    \caption{
    Using as an example GW$190413\_052954$, we plot the PDF of primary mass $M_1= 33.7 \, M_{\odot}$ (shown as a grey line). The primary mass, within $90\%$ credible interval, has an upper limit $M_{1}^{\textrm{upper}}=33.7 + 10.4 = 44.1 \, M_{\odot}$ and a lower mass limit $M_{1}^{\textrm{lower}}=33.7 - 6.4 = 27.3 \, M_{\odot}$ (light orange dashed lines).}
    \label{fig:GW_PDF}
\end{figure}

For each event, we search for the optimal combination of parameters $\xi$, $\omega$ and $\alpha$ that fits the relevant PDF of $M_1$~\ref{fig:GW_PDF}. 
We also obtain the optimal combination of $\xi$, $\omega$ and $\alpha$ parameters for $M_2$ in the same way. 
In Fig.~\ref{fig:GW_PDF}, we show as an example, the estimated PDF for event GW$190413\_052954$ with a primary mass $M_1= 33.7 \, M_{\odot}$.

To obtain the PDF for $q$ of each event, we pick a random value from the event's distribution of $M_2$ and divide by a random value from the event's distribution of $M_1$. Since $q$ is by definition the ratio between the less massive event and the more massive event, in the case where the value randomly picked from the $M_1$ distribution ($M_{1,\textrm{expl}}$), is smaller than the randomly picked value from the $M_2$ distribution($M_{2,\textrm{expl}}$), i.e $M_{1,\textrm{expl}}<M_{2,\textrm{expl}}$, the division yields a value of $q$ that is larger than 1. We pick in this case $1/q$ instead.
We repeat this process a thousand times to obtain the PDF for q for that event; and proceed in the same way to obtain the PDF for $q$ for all events.

\subsection{Mass-distribution properties of the observed binary black holes}

\begin{figure*}[!th]
    \centering
    \includegraphics[width=0.9\textwidth]{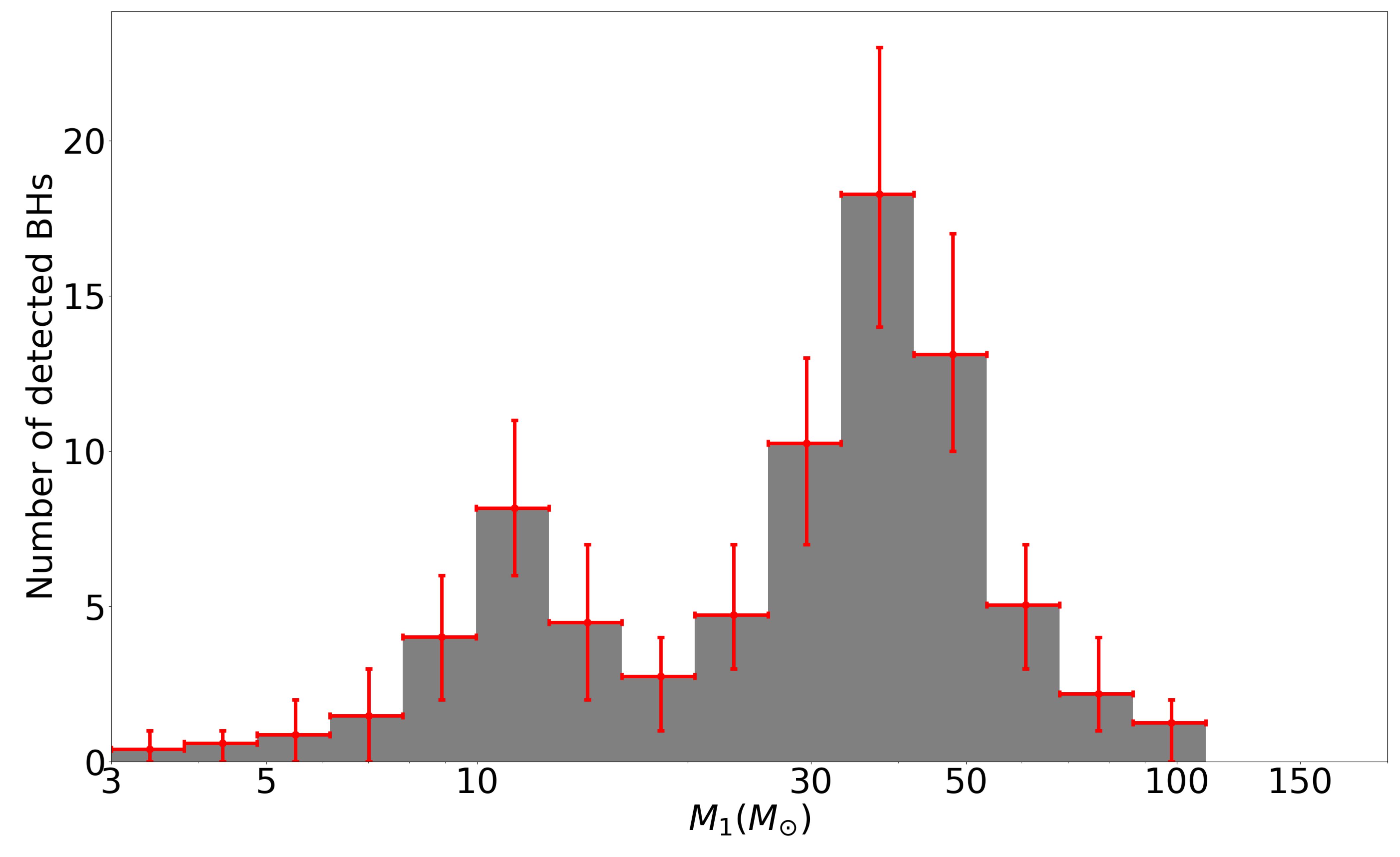}
    \caption{Histogram of $M_1$ mass distribution for BBHs, with Poisson error bars from GWTC3 events.}
    \label{fig:M1hist}
\end{figure*}

With the skew-normal parameters for the quantities $M_1$ and $M_2$, we have access to the PDFs of $M_1$ and $q$  for all events. For each event, we make a histogram from the PDF of $M_1$, using 15 discrete bins. We give our $M_{1}$ mass-binning in Table~\ref{table:binning}. 
We record the number of counts per bin ($n_{i,j}$ with i indexing the events and j the bins). To obtain the mass distribution of the $M_1$ parameter for the observed LVK BBHs, we make a histogram with bin counts $n_{tot,j}$ (with $j$ the index of the bin) equal to the sum of the bin counts from the histograms of each event,
\begin{equation}
    n_{\textrm{tot},j}= \sum_i n_{i,j}.
\end{equation}
The bin counts of our final histogram for the $M_1$ distribution is then $n_{\textrm{tot},j}$ for bin $j$.
In Fig.~\ref{fig:M1hist}, we provide the derived $M_{1}$-histogram for the 72 events with FAR of $< 1$ $\textrm{yr}^{-1}$. 

\begin{table}[h!]
\centering
\begin{tabular}{|c | c | c | c|} 
 \hline
 Quantity & Number of bins & log or lin space & range of values \\ [0.5ex] 
 \hline
 $M_1$ & 15 & logarithmic & $3M_\odot$ to $200M_\odot$\\ 
 \hline
 $q$ & 10 & linear & 0 to 1\\
 \hline
\end{tabular}
\caption{The parametric binning of the histograms for $M_1$ and $q$.}
\label{table:binning}
\end{table}

\begin{figure*}
    \centering
    \includegraphics[width=0.9\textwidth]{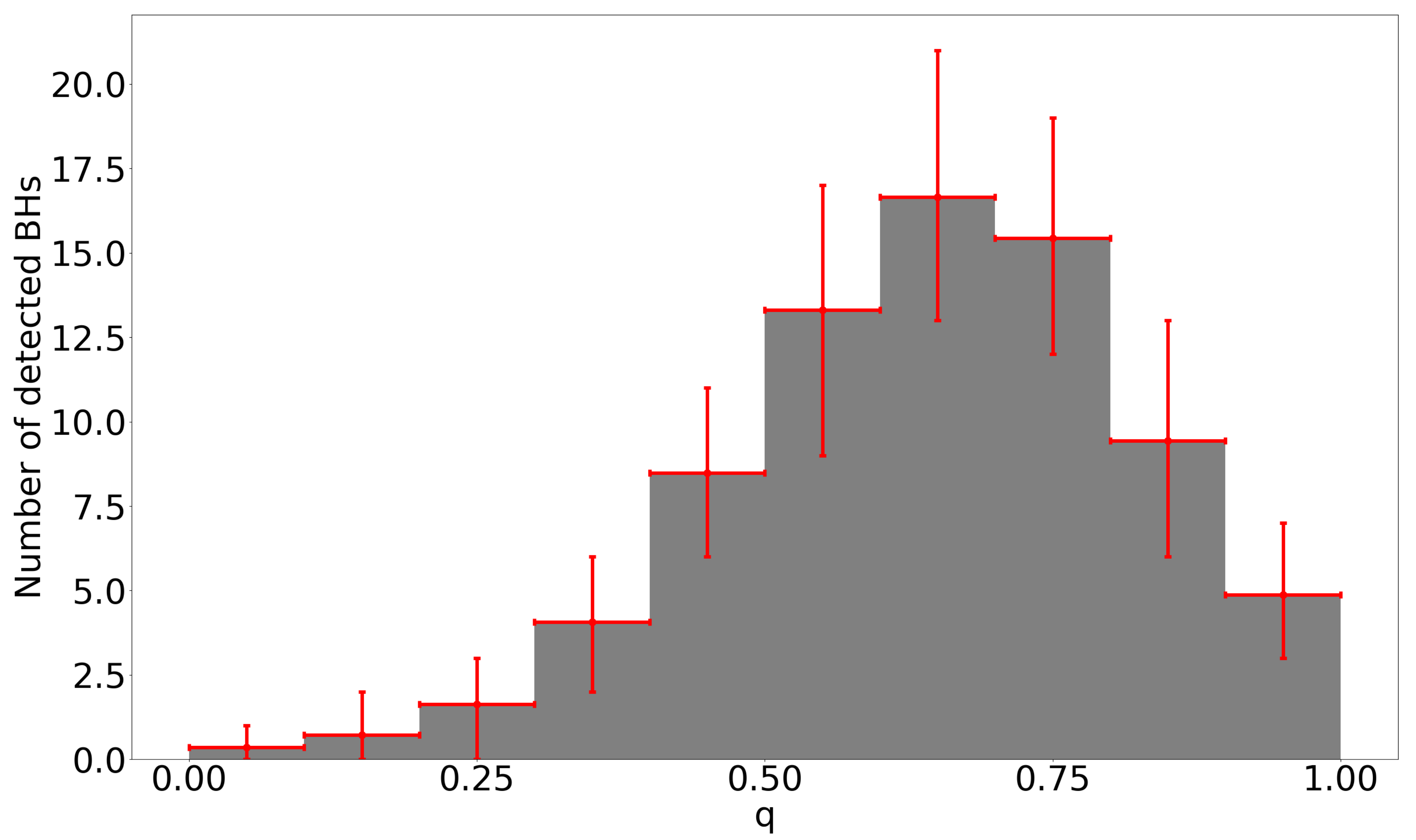}
    \caption{Histogram of the $q$-distribution of BBHs, with Poisson error bars from GWTC3 events.}
    \label{fig:qhist}
\end{figure*}

We proceed in the same way starting from the PDFs of $q$ for each event to obtain the histogram for the $q$ distribution of the 72 BBH mergers, the $q$-binning is also given in Table~\ref{table:binning} and shown in Fig.~\ref{fig:qhist}.

Since we have a small number of events per bin from the O1, O2 and O3 runs, we use Poisson distribution to obtain the error bars for these histograms. The $1\sigma$-error for each bin in the mass distribution of BBHs is the $1\sigma$-error of the Poisson distribution drawn from the number of bin counts in the bin considered. This method is valid for the histograms of both $M_1$ and $q$.

\section{Simulated BBHs and the $\chi^2$ test}
\label{sec:Simulated_BBHs}

To reach a better understanding of the mass distribution of the BHs detected by the LVK collaboration, we sample BHs following a power-law and  a power-law$+$Gaussian peak (PP) mass-distribution. We place the BHs at a redshift following a power-law distribution for their comoving rate density, 
\begin{equation}
    \frac{dR}{d(z+1)} \propto (1+z)^\kappa.
\end{equation}

Ref.~\cite{PhysRevX.13.011048}, used the same parameterization for the redshift distribution of BBHs and found a best fit range for  $\kappa=2.9^{+1.7}_{-1.8}$. We use the central value of $\kappa = 2.9$. Since the observed event with the largest redshift has a redshift with a central value of 1.18, we simulate events with a redshift of up to 1.5.
Then, on the simulated events, we
calculate the SNR observed by the LIGO interferometers (discussed in further detail in section~\ref{subsec:SNR}). 
We select the simulated events with a SNR larger than 8 and normalise the distribution of detectable events to perform our $\chi^2$ test.

\subsection{The BBH properties' distributions}
\label{subsec:BBH prop dist}

We consider two mass distributions from which we will sample BBHs. We start with the power-law distribution, in which the primary mass $M_1$ and $q$ are drawn from,

\begin{equation}
    \frac{dN}{dM_1} \propto M_1^{-\alpha}.
\end{equation}

\begin{equation}
    \frac{dN}{dq} \propto q^\beta.
\end{equation}
Where $\alpha$ is the exponent of the power-law distribution from which we draw $M_1$ and $\beta$ the exponent of the power-law distribution from which we draw $q$.
We consider the power-law distribution as a simplified mass distribution of astrophysical BHs.

The second distribution we consider is a PP distribution, which is a power-law distribution with an additional Gaussian component. 
The parameters are the exponent of the distributions for $M_1$ and $q$, i.e. $\alpha$ and $\beta$ and the location parameter of the Gaussian peak $\mu$. We are using the Gaussian in this distribution to simulate monochromatic PBHs or BHs from earlier mergers. 
Since PBHs would be indistinguishable from astrophysical black holes, we pick for the width of the Gaussian $\sigma$, to be the average error of the primary and secondary BH masses detected by LVK in an interval centered at the considered peak location. 
This approximation consists in averaging the errors on the detected BH masses at different redshifts. This is a simple way to estimate the LIGO detectors' sensitivity accounting for how that sensitivity changes with respect to redshift.
Given the small sample of BBH detections, we consider this an adequate approximation, leaving a more sophisticated approach for future comparisons to observations with many more detections.   
We note that in order to have a better estimate of the mass sensitivity of the LIGO detectors in this mass range, both primary and secondary masses in the relevant mass interval are used.
Our approximation allows to reduce the number of parameters probed for PP distribution, as the width of the Gaussian becomes dependant on the location of the peak.  

Using the PP model, we simulate events from the power-law component and events from the Gaussian component independently. If that component is associated to PBHs, this assumes that there is no term in the observed events from merging binaries consisting of a PBH and a regular BH from stellar evolution.
Such binaries are expected to be very rare.
PBHs occupy the entire dark matter halo. 
The PBH-PBH merger events occur at the smaller mass dark matter halos and at volumes within the halos that is not occupied by much regular baryonic mass and thus stars \cite{Bird:2016dcv, Sasaki:2016jop, Ali-Haimoud:2017rtz, Aljaf:2024fru}.  Only a small fraction of PBHs go through the inner part of a galaxy where BH from stellar evolution may lay. In fact, the PBHs have relatively high velocities as they cross the inner galaxies, suppressing their direct capture cross section by other BHs. 

\subsection{Evaluating the Signal to Noise Ratio}
\label{subsec:SNR}

\begin{figure}
    \centering
    \vspace{-0.2in}
    \includegraphics[width=0.53\textwidth]{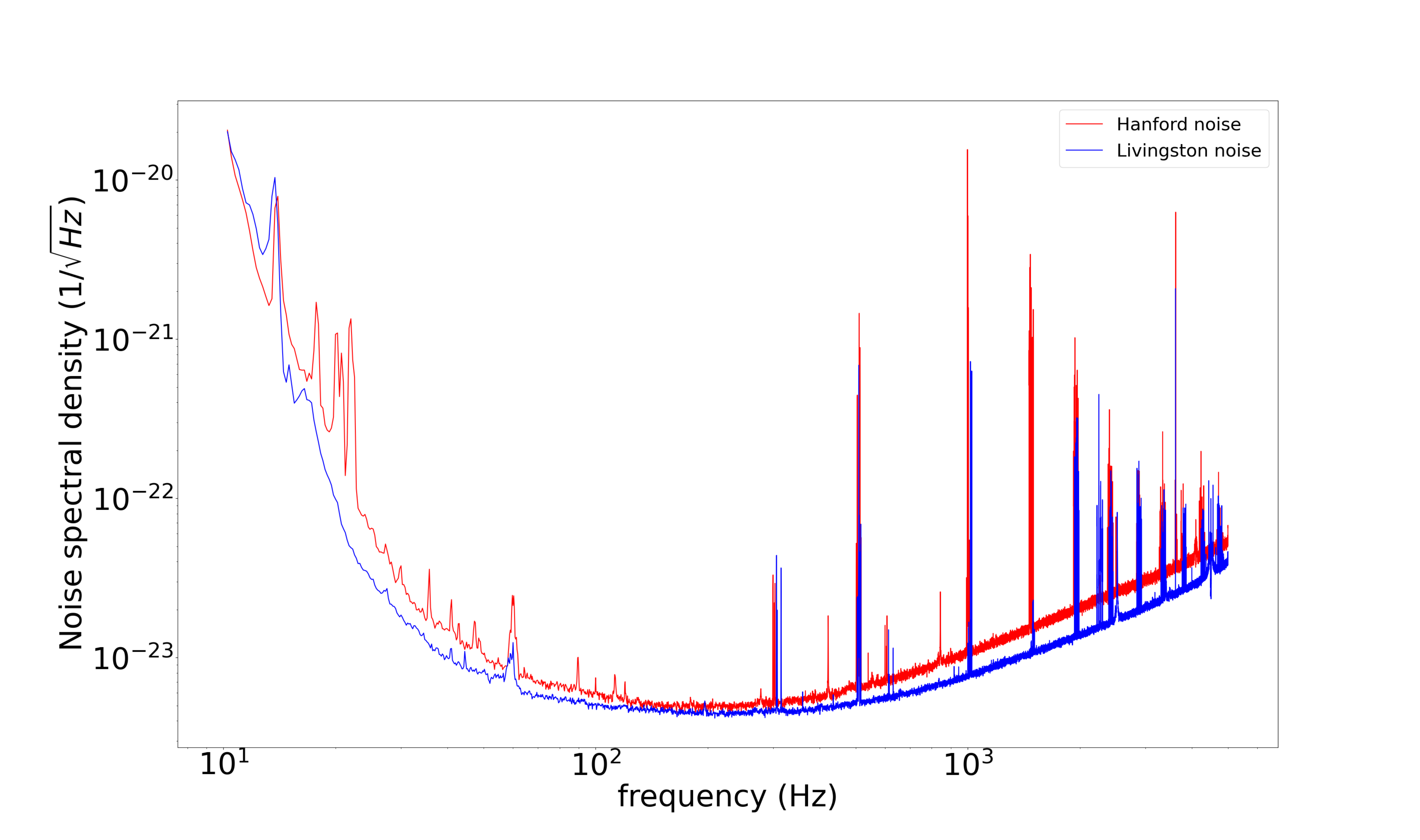}
    \caption{The noise curves of LIGO Hanford and LIGO Livingston used in this work, taken from \cite{LIGOnoise}.}
    \label{fig:HLnoise}
\end{figure}

After simulating BBHs and placing them at a certain redshift, we need to calculate the SNR of the simulated events observed by LVK.
Since the two LIGO interferometers are significantly more sensitive than the Virgo or KAGRA ones, we evaluate the SNR from the just the LIGO observatory.
We use the LIGO noise curves of Ref.~\cite{LIGOnoise}, re-plotted in Fig.~\ref{fig:HLnoise} for reference. Ref.~\cite{LIGOnoise}, provides
one noise curve for the Hanford site and one noise curve for the site in Livingston based on the first three months of the O3 run. 
These noise curves do not account for the change in noise with respect to time. We have noticed that this leads to typically less massive events being detectable in our simulations than claimed in the small sample of LVK collaboration. We leave is point to be further discussed in our Results section~\ref{sec:results}.

To calculate the SNR we need the expressions of the merger and quasinormal (qnr) mode frequencies as well as the expression of the energy density spectra of the inspiral, merger and ringdown phases. All these expressions are in geometrised units ($G=c=1$), \cite{PhysRevD.57.4535}.
The merger frequency is given by,
\begin{equation}
    f_m = \frac{0.02}{M}.
\end{equation}
The qnr frequency is,
\begin{equation}
    f_{\textrm{qnr}}=\frac{0.13}{M}.
\end{equation}

The energy density spectrum of the inspiral phase for a frequency lower than $f_m$ is given by,
\begin{equation}
    \frac{dE}{df}_i(f) = \frac{1}{3}\pi^{2/3}\mu M^{2/3}f^{-1/3},
    \label{eq:inspiral}
\end{equation}
evaluated at the source.
The energy density spectrum of the merger phase for a frequency between $f_m$ and $f_{\textrm{qnr}}$ is given by,
\begin{equation}
    \frac{dE}{df}_m(f) = 0.91M^2 F(\mu/M).
\end{equation}
The energy density spectrum of the ringdown phase for a frequency around $f_{\textrm{qnr}}$ is given by,
\begin{equation}
    \frac{dE}{df}_r(f)= \frac{1}{8}A^2 \cdot Q \cdot M^2 f_{\textrm{qnr}}.
\end{equation}
With $M$ the total mass: $M=M_1 + M_2$ , $\mu$ the reduced mass : $\mu = \frac{M_1 M_2}{M_1 + M_2}$ and $F(\mu/M)=(\frac{4\mu}{M})^2$. We pick A=0.4 and $Q=2(1-a)^{\frac{-9}{20}}$ with a=0.67 for the ringdown energy density spectrum.

With the expressions on the energy spectra of the phases of BBHs coalescence, we can express the characteristic amplitude of the signal,
\begin{equation}
    h_c^2 = 2 \frac{(1+z)^2}{(\pi D_L(z))^2} \frac{dE}{df}((1+z)f).
\end{equation}
With $z$ the redshift at which the BBH is placed, $\frac{dE}{df}$ is the energy spectrum of the coalescence, which is the sum of the energy spectra of the inspiral, merger and ringdown phases: $\frac{dE}{df} = \frac{dE}{df}_i + \frac{dE}{df}_m + \frac{dE}{df}_r$. And $D_L(z)$ is the luminosity distance of the BBH at a regshift z,
\begin{equation}
D_L(z) = (1+z)\int \frac{c}{H_0\sqrt{\Omega_m(1+z)^3+\Omega_\Lambda}}dz,
\end{equation}
where c is the speed of light, $H_0$ the Hubble constant, $\Omega_m$ the matter density parameter and $\Omega_\Lambda$ the dark energy density parameter obtained from "the best fit Euclidian $\Lambda$CDM cosmology from Plank Collaboration(2018b)" according to \cite{dodelson:2020} \cite{Planck:2018vyg}. 
Since $h_c$ is measured on Earth we need to account for the redshift effect on the frequency, we thus take the energy spectum at $(1+z)f$.
From \cite{LIGOnoise}, we obtain a strain noise for the Livingston detector that we note $h_{n,L}$ and a strain noise for the Hanford detector that we note $h_{n,H}$.

To evaluate the SNR $\rho$ we combine them as,
\begin{equation}
    <\rho> = \sqrt{\frac{4}{5}\int \frac{h_c / (2f)^2}{h_{n,L}^2} + \frac{4}{5}\int \frac{h_c / (2f)^2}{h_{n,H}^2}}.
    \label{eq:SNR}
\end{equation}
We use the two noise curves separately in the expression of the SNR because the Livingston noise we use is consistently lower than the Hanford one (see Ref.~\ref{fig:HLnoise}). Using the two noise curves independently in the expression of the SNR instead of an average of two curves does not change the value of the SNR significantly. Nonetheless,  the expression of Eq.~\ref{eq:SNR} is more accurate.

We simulate $10^{6}$ BBH with masses from 4 to 150 $M_{\odot}$ and up to a redshift of 1.5. 
Since most generated binaries will have large luminosity distances and thus a SNR$<8$, we make a first estimate to disregard binaries for which our SNR will be well bellow 8. This allows us to optimise significantly our computing time of SNR for the remaining sample of merging binaries. 
In simple terms, we need to evaluate the maximum redshift at which a binary described by $M_{1}$ and $q$ would give a SNR$\geq8$. We note that we do account for the inclination angle as this is not included in Eq.~\ref{eq:inspiral}.
Discarding detactable BBHs in the study would result in potentially deforming the distributions from which we draw the quantities that make our simulated BBHs. This is why, for each range of primary masses $M_1$, we pick a cutoff value for the redshift z that is slightly larger than the found maximum redshifts. This is to ensure that all detectable simulated BBHs are included in the study. In Table~\ref{table:MaxRedshift}, we show the largest redhsift for specific choices of $M_{1}$. For low $M_{1}$ values, this translates to setting $q=1$. However, for very high $M_{1}$ values, the low-frequency noise curve prevents the detection of binaries with $q\simeq 1$. 

\begin{table}[h!]
\centering
\begin{tabular}{| c | c | c|} 
 \hline
  $M_1$ ($M_\odot$) & highest value of redshift & cutoff value for $z$ \\ [0.5ex] 
 \hline
 6 & 0.09906 & 0.11 \\ 
 \hline
 10 & 0.12489 & 0.13 \\
 \hline
 15 & 0.15210 & 0.17 \\  
 \hline
 25 & 0.20155 & 0.22 \\ 
 \hline
 90 & 0.60272 & 0.7 \\ 
 \hline
\end{tabular}
\caption{The largest redshift at which a binary with primary mass $M_1$, would be observed with SNR$>8$. We test the range of possible values for $q$. We simulate $10^{6}$ BBHs and present our result in the second column. The third column gives cutoff value of $z$ that we pick for that $M_1$ value, in selecting binaries (see text for details).}
\label{table:MaxRedshift}
\end{table}

\subsection{Testing models with data}

We calculate the SNR of the simulated BBHs that LIGO would detect and save the primary mass, the ratio $q$ and the redshift of the simulated events with an SNR$>8$. These events constitute our simulated model. We make a histogram of $M_{1}$ and $q$ distributions of the simulated model keeping the same number of bins that we use for the LVK data with the same range for $M_{1}$ and $q$.

One of our objectives, is to find the minimal $\chi^2$ for each choice of parameters. This minimisation is done by searching for the normalisation factors that minimise the $\chi^2$ value. For the power-law model, one normalisation factor is enough to minimise the $\chi^2$ value. The candidates for the minimal $\chi^2$ used to compare the $M_1$ distribution of the simulated events for one value of $\alpha$ and one value of $\beta$ with the $M_1$ distribution of the LVK data have the following expression,
\begin{equation}
    \chi^2_{M_1} = \sum_{i=1}^{15}(\frac{data_{M_1,i} - N \cdot model_{M_1,i}}{sigma_{M_1,i}})^2.
    \label{eq:chi_sqr_M_1}
\end{equation}
Where, $i$ labels the bins and goes from 1 to 15 for $M_1$, $data_{M_1,i}$ is the number of $M_1$ counts from the LVK data in the bin $i$, $model_{M_1,i}$ is the number of $M_1$ counts from the simulated events in the bin $i$, $N$ a normalisation factor, and $sigma_{M_1,i}$ is the 1$\sigma$ error on $data_{M_1,i}$.

Similarly, the candidates for the minimal $\chi^2$ for the $q$ distribution have the following expression,
\begin{equation}
    \chi^2_{q} = \sum_{i=1}^{10}(\frac{data_{q,i} - N \cdot model_{q,i}}{sigma_{q,i}})^2.
    \label{eq:chi_sqr_q}
\end{equation}
Again, $i$ labels the bins and goes from 1 to 10 for $q$, $data_{q,i}$ is the number of $q$ counts from the LVK data in the bin $i$, $model_{q,i}$ is the number of $q$ counts from the simulated events in the bin $i$, $N$ a normalisation factor, and $sigma_{q,i}$ is the 1$\sigma$ error on $data_{q,i}$.\\

The total $\chi^2$ is then,
\begin{equation}
    \chi^2_{tot} = \chi^2_{M_1} + \chi^2_q.
    \label{eq:chi_sqr_tot}
\end{equation}

For the PP model, we have two components for the distribution, a component following a power-law distribution and a component following a Gaussian distribution. The expression for $\chi^2_{M_1}$ and $\chi^2_{q}$ that is very similar to equations \ref{eq:chi_sqr_M_1} and \ref{eq:chi_sqr_q},
\begin{equation}
    \chi^2_{M_1,PP} = \sum_{i=1}^{15}(\frac{data_{M_1,i} - model_{M_{1,PP},i}}{sigma_{M_1,i}})^2 ,   
\end{equation}
where,
\begin{eqnarray}
model_{M_{1,PP},i} &=& N_{PL}\cdot model_{M_{1,PL},i} \nonumber \\
&+& N_{Peak} \cdot model_{M_{1,Peak},i}.
\end{eqnarray}

Similarly,
\begin{equation}
        \chi^2_{q,PP} = \sum_{i=1}^{10}(\frac{data_{q,i} - model_{q_{PP},i}}{sigma_{q,i}})^2
\end{equation}
with,
\begin{eqnarray}
model_{q_{PP},i} &=& N_{PL}\cdot model_{q_{PL},i} \nonumber \\
&+& N_{Peak} \cdot model_{q_{Peak},i}.
\end{eqnarray}
In this case we have one normalization factor for the power-law component, indexed with "PL", and one for the Gaussian component, indexed with "Peak". Bin counts from the simulated events also come from two different components of the distribution, the same indices introduced on the normalization factors apply here.
The total $\chi^2$ result  of the PP model is again given by Eq.~\ref{eq:chi_sqr_tot} and  minimised to find the best-fit combination of  normalization factors $N_{PL}$ and $N_{Peak}$.

Note that having two different normalization factors for the power-law component and the Gaussian components of the PP model is consistent with what the two components represent; since the BHs from stellar
evolution and the PBHs have different origins and different dynamics \footnote{For the case where the Gaussian component represents BHs from earlier BBH mergers, still the relative normalizations of the PL and the Peak components may be weakly correlated.}.

\section{Results} 
\label{sec:results}

\subsection{Power-law model}
\label{subsec:PL_results}
For the single power-law model, from the $\chi^2$ test we find the best fit values of $\alpha = 3.44^{+0.2}_{-0.29}$ and $\beta = -0.89^{+0.5}_{-0.1}$. These give a total $\chi^2$ of $\chi^2_{tot}=38.04$. 
The quoted confidence intervals for the best fit parameters are $90\%$. The $\chi^2$ minimization was done with one parameter (the normalization factor for the power-law model), leaving us with 24 degrees of freedom for this model. 
The $\chi^2$ per degree of freedom ($\chi^2_{\textrm{d.o.f.}}$) in this case is then $\chi^2_{\textrm{d.o.f.}} = 1.58$. 
Table~\ref{table:chi_sqr_PL_results}, includes this result along with $\chi^2$ values for specific parameter choices as fixing $\beta$ to 0 and 1 or $\alpha$ to 2.3 and reporting the best-fit value for the other parameter. 

We find values for $\alpha$  that are consistent with Ref.~\cite{PhysRevX.13.011048}. The values we find for $\beta$ however are lower than those reported by Ref.~\cite{PhysRevX.13.011048}, i.e. preference for smaller values of $q$. 

Given the quoted uncertainties of Ref.~\cite{PhysRevX.13.011048}, our best fit value for $\beta$ is beyond their 90$\%$ confidence interval. That is not the case for the $\alpha$ parameter however.
In our opinion this suggests that this method of analyzing the gravitational wave merger data is somewhat orthogonal to that of Ref.~\cite{PhysRevX.13.011048}. With a better understanding of the LIGO noise-curve evolution, the differences between the two methods should become less pronounced.



\begin{figure*}[!th]
    \centering
    \includegraphics[width=0.8\textwidth]{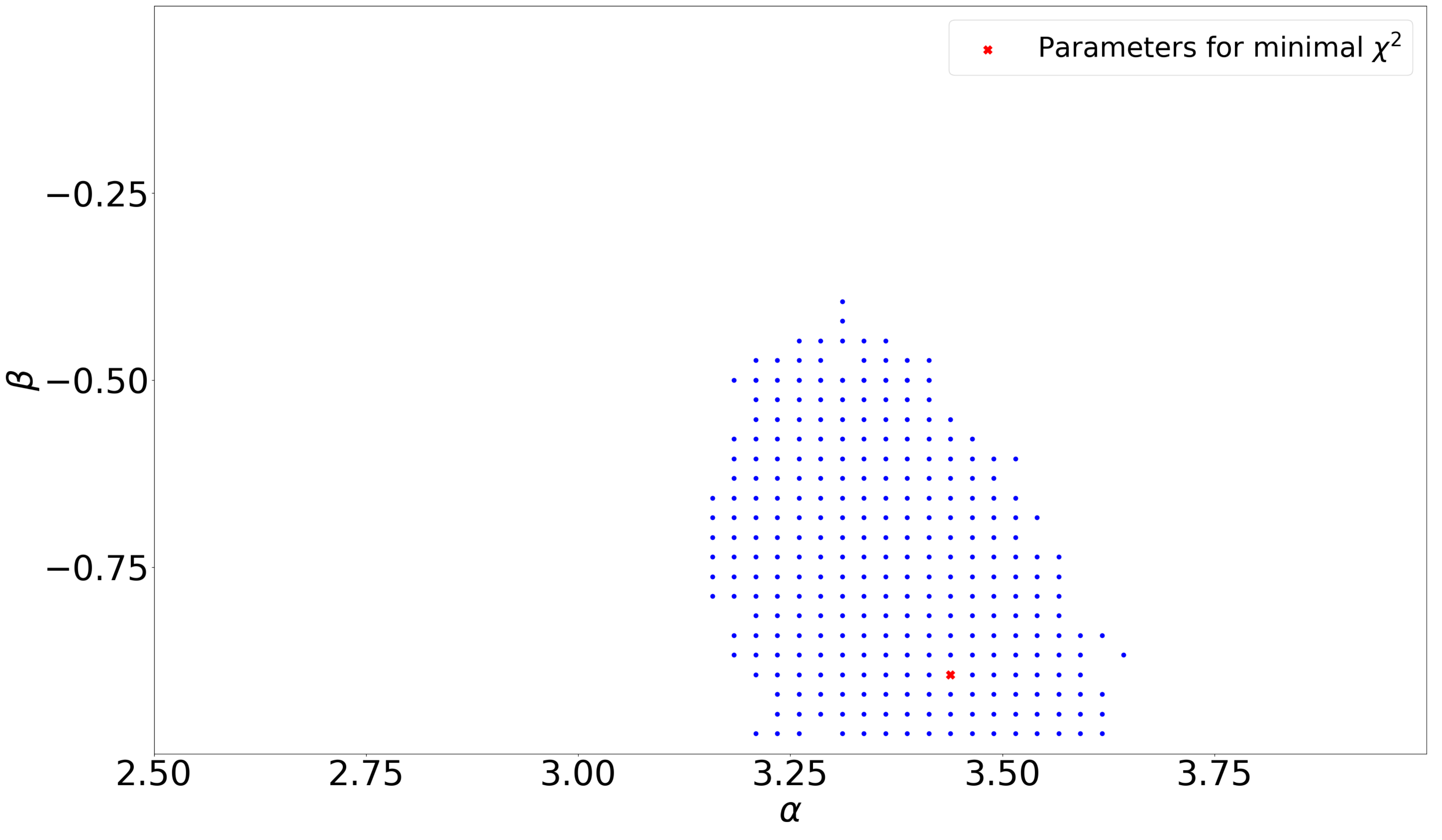}
    \caption{The $\alpha$-$\beta$ parameter space for the power-law model probed by the LVK BBHs. The blue dots are inside the $90\%$ credible interval around the point giving the minimum $\chi^2$ value. (red ``x'').}
    \label{fig:PL space param good chi}
\end{figure*}

\begin{figure*}[!th]
    \centering
    \includegraphics[width=0.8\textwidth]{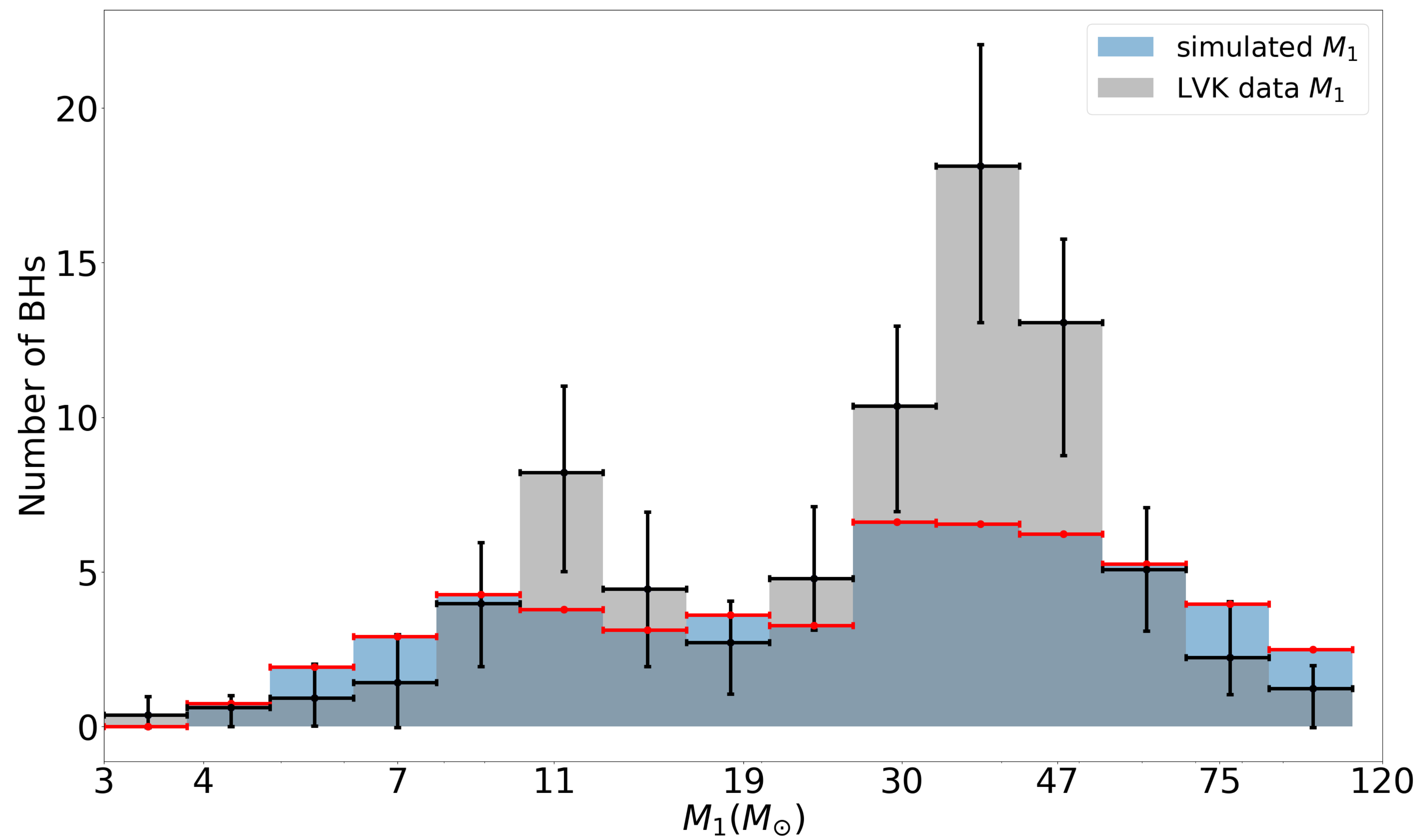}
    \caption{Normalised histogram of the distribution of $M_1$, of simulated BBHs detectable with a SNR$>8$. The underlying simulated distribution is that of a power-law model with $\alpha = 3.44$, and $\beta = -0.89$ (blue columns with red best-fit normalizations). Superposed to that, is the histogram of the $M_1$ distribution from the LVK BBHs with Poisson error-bars (grey columns with back errors).}
    \label{fig:opt_M_1_PL}
\end{figure*}

\begin{figure*}[!th]
    \centering
    \includegraphics[width=0.8\textwidth]{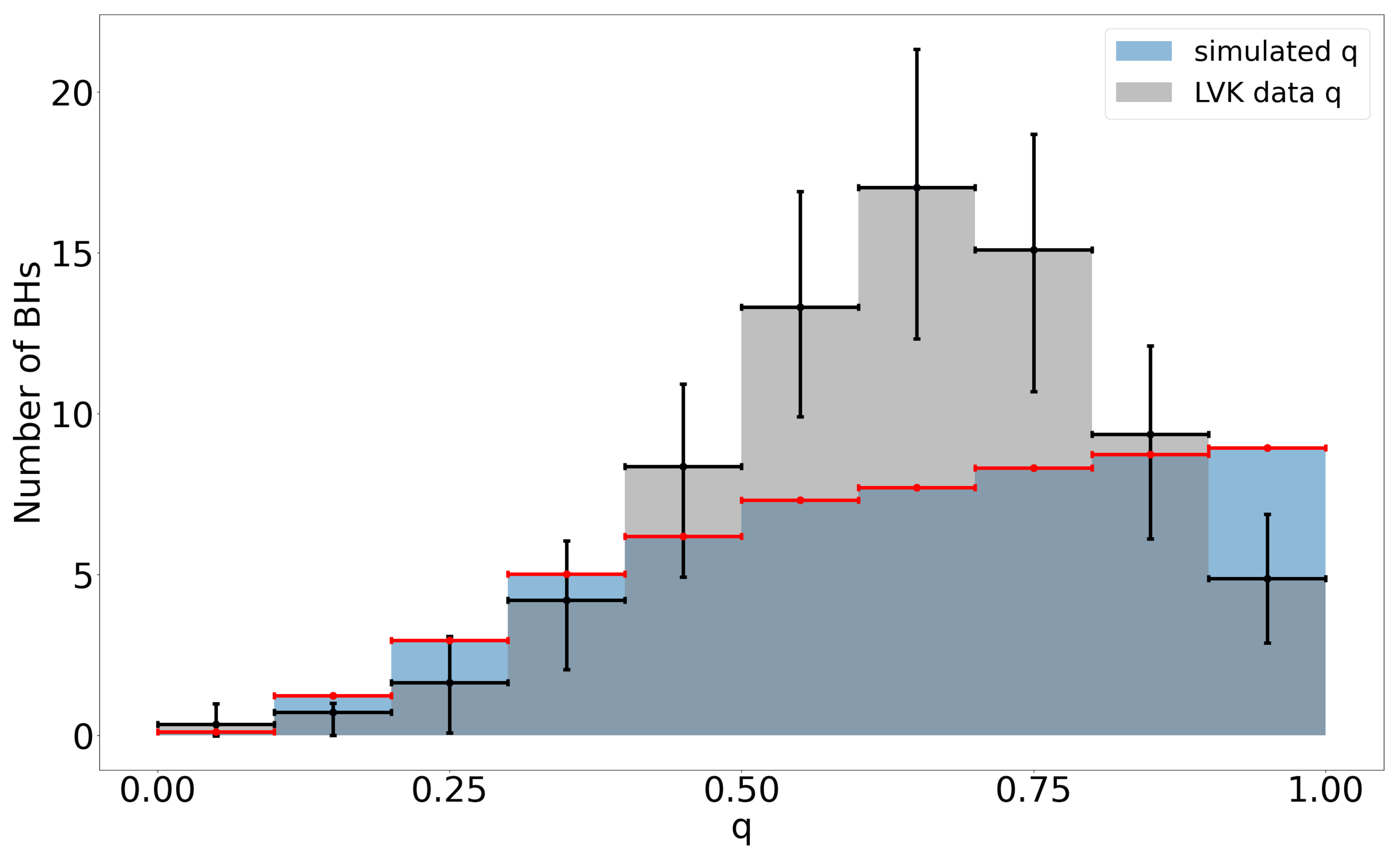}
    \caption{Normalized histogram of the distribution of $q$ of simulated BBHs, detectable with an SNR$>8$. In our simulation the BBHs follow a power-law model with $\alpha = 3.44$, and $\beta = -0.89$ (blue columns with red normalizations). As in Fig.~\ref{fig:opt_M_1_PL}, superposed is the $q$ distribution from the LVK BBHs (grey histogram with black Poisson errors).}
    \label{fig:opt_q_PL}
\end{figure*}

In Fig.~\ref{fig:PL space param good chi}, we show the area in parameter space around the best-fit values that is within $90\%$ confidence interval.
The best-fit point is given by the red ``x'' symbol. 
In Figs.~\ref{fig:opt_M_1_PL} and \ref{fig:opt_q_PL}, we show the $M_{1}$ and $q$ histograms of the simulated BBHs from a power-law model with the best-fit parameters. These are shown as blue histograms with red  best-fit normalization values (no errors shown). For comparison we include the histograms constructed from the LVK data (grey histogram with black Poisson errors). 

\begin{table}[h!]
\centering
\begin{tabular}{|c | c | c | c | c|} 
 \hline
 Value of $\alpha$ & Value of $\beta$ & Best $\chi^2_{M_1}$ & best $\chi^2_q$ & Best $\chi^2_{\textrm{d.o.f.}}$ \\ [0.5ex] 
 \hline
 3.44 & -0.89 & 22.09 & 17.75 & 1.58\\ 
 \hline
 3.23 & 0 & 23.47 & 25.7 & 2.05\\ [1ex] 
 \hline
  3.16 & 1 & 27.52 & 40.46 & 2.83\\ [1ex] 
 \hline
  2.3 & -0.45 & 55.22 & 22.21 & 3.23\\ [1ex] 
 \hline
\end{tabular}
\caption{Results of the $\chi^2$ test after comparing simulated populations from the power-law model to the 
LVK data. The number of degrees of freedom for this test is 24. The first line shows the best fit parameters and $\chi^2$ results, the second line the best fit results for $\beta=0$, the sthird line the best fit results for $\beta=1$ and the last line the best fit results for $\alpha = 2.3$, which is the power-law exponent for the initial mass function of stars with masses > $3M_\odot$   Ref.~\cite{Kroupa_2001}.}
\label{table:chi_sqr_PL_results}
\end{table} 

As can be seen from Fig.~\ref{fig:opt_M_1_PL}, the 
single power-law can not explain the peak around 40 $M_{\odot}$ on the $M_{1}$-distribution.
This is also clear from the best-fit  $\chi^2_{\textrm{d.o.f}}$ result for the power-law model. It is significantly larger than one. 
The power-law model, is not sufficient to describe the mass distribution of the LVK collaboration's events. It is important to note that when working with the power-law model, we are effectively considering mergers between first-generation BHs. Accounting for hierarchical mergers may require a more complex model-distribution and lead to different results. In fact, our result suggest the need for some additional component to the simple power-law model, which at a first level we address in Section~\ref{subsec:PP_results}.

The results obtained in this section relied on calculating the SNR from simulated mergers that would be detected with by LIGO. The obtained SNR is dependent on the noise curve that we use for the two LIGO interferometers. 
We have access to only one noise curve for the LIGO Hanford and one for the LIGO Livingston sites \cite{LIGOnoise}.
LIGO's noise is time dependent and keeps changing as the detectors are undergoing improvements. 
While testing our code to calculate the SNRs, we compared the SNR we obtained for specific LVK's events with their reported SNR. 
We noticed that some events with large masses where detected by LVK with a SNR larger than 8, while we obtained a SNR smaller than 8. 
This can be explained by a seismic noise at the time of observation that was smaller than the seismic noise accounted for in the noise curves we have available. 
Having access to noise curves for the Hanford and Livingston for different times during the observation period would solve this problem. 

\subsection{Power-law + Peak model}
\label{subsec:PP_results}
As discussed in \ref{subsec:BBH prop dist}, the central value $\mu$ of the Gaussian distribution has a corresponding width $\sigma$, that its value is correlated to the value of $\mu$. 
To obtain $\sigma$, we average the $90\%$ errors in the LVK catalog from all $M_1$s and $M_2$s,  each time within an interval spanning from $\mu - 5M_\odot$ to $\mu + 5M_\odot$. 
The derived result is then divided by 2 and converted into the $1\sigma$ error. 
We test 12 discrete choices for the peak value of the Gaussian distribution $\mu$.
The corresponding standard deviations are given in Table\ref{table:sigma_of_peak}, for each choice of $\mu$. 

\begin{table}[ht!]
\centering
\begin{tabular}{| c | c |} 
 \hline
 Peak value ($\mu$) & Corresponding $1\sigma$ \\ [0.5ex] 
 \hline
 5$M_\odot$ & 0.9447 \\ 
 \hline
 10$M_\odot$ & 2.086 \\
  \hline
 15$M_\odot$ & 2.873 \\
  \hline
 20$M_\odot$ & 4.134 \\
  \hline
 25$M_\odot$ & 4.581 \\
  \hline
 30$M_\odot$ & 5.091 \\
  \hline
 35$M_\odot$ & 5.683 \\
  \hline
 40$M_\odot$ & 6.380 \\
  \hline
 45$M_\odot$ & 7.094 \\
  \hline
 50$M_\odot$ & 12.5 \\
  \hline
 60$M_\odot$ & 14.3 \\
  \hline
 80$M_\odot$ & 19.1 \\
 \hline
\end{tabular}
\caption{The location parameter values probed for Gaussian component of the PP model and their corresponding standard deviations (see text for details).}
\label{table:sigma_of_peak}
\end{table}

The best fit values for the $\chi^2$ test for the PP model are achieved for $\alpha = 3.34^{+0.36}_{-0.15}$, $\beta=-1^{+0.37}_{-0}$ and a Gaussian peak at $40M_\odot$. For that value of $\mu$, the corresponding standard deviation is $6.38M_\odot$. We get a $\chi^2_{M_1}=16.74$, $\chi^2_q=16.05$ and a $\chi^2_{\textrm{tot}} = 32.79$. In this case, the $\chi^2$ per degree of freedom is $\chi^2_{\textrm{d.o.f.}} = 1.42$. This is a significant improvement to the simple PL model's best fit results.

With the quoted best-fit choices, about one in five of simulated events are drawn from the Gaussian component. 
We note that the $\chi^2$ from the fitting of $M_1$ alone, gives a result very close to 1 per d.o.f. .
For the PP case, it is the $q$-distribution that contributes the most to the $\chi^2_{\textrm{tot}}$. 
This seems to suggest that modeling the BBH population as either a simple power-law for both $M_{1}$ and $q$, or that in combination with a Gaussian component for $M_1$, is not sufficient to explain the observed $q$-distribution.
To explain the BBH observations, we require a mechanism that favours values of around 0.6. A power-law for $q$, will consistently lead to underestimating the number of counts in the the $q$-bin centred at 0.6, or overestimate the the number of counts in the last bins.

Table~\ref{table:chi_sqr_PP_results}, includes our best-fit choices for the PP model, along with $\chi^2$ values for specific parameter choices as fixing $\beta$ to 0 and 1 or $\alpha$ to 2.3 and marginalizing over the other parameters. We report the $\chi^{2}$ values and the best-fit values for the other free parameters. Ref.~\cite{PhysRevX.13.011048}, found evidence for a peak with a best-fit value of 45 $M_{\odot}$, with an $\alpha$ of 3.34, while Ref.~\cite{ulrich2024multigenerationalblackholepopulation} found the best-fit value of the peak to be at 35 $M_{\odot}$. Those are in agreement to our result for a best-fit value of a peak at 40 $M_{\odot}$. 

\begin{figure*}[!th]
    \centering
    \includegraphics[width=0.8\textwidth]{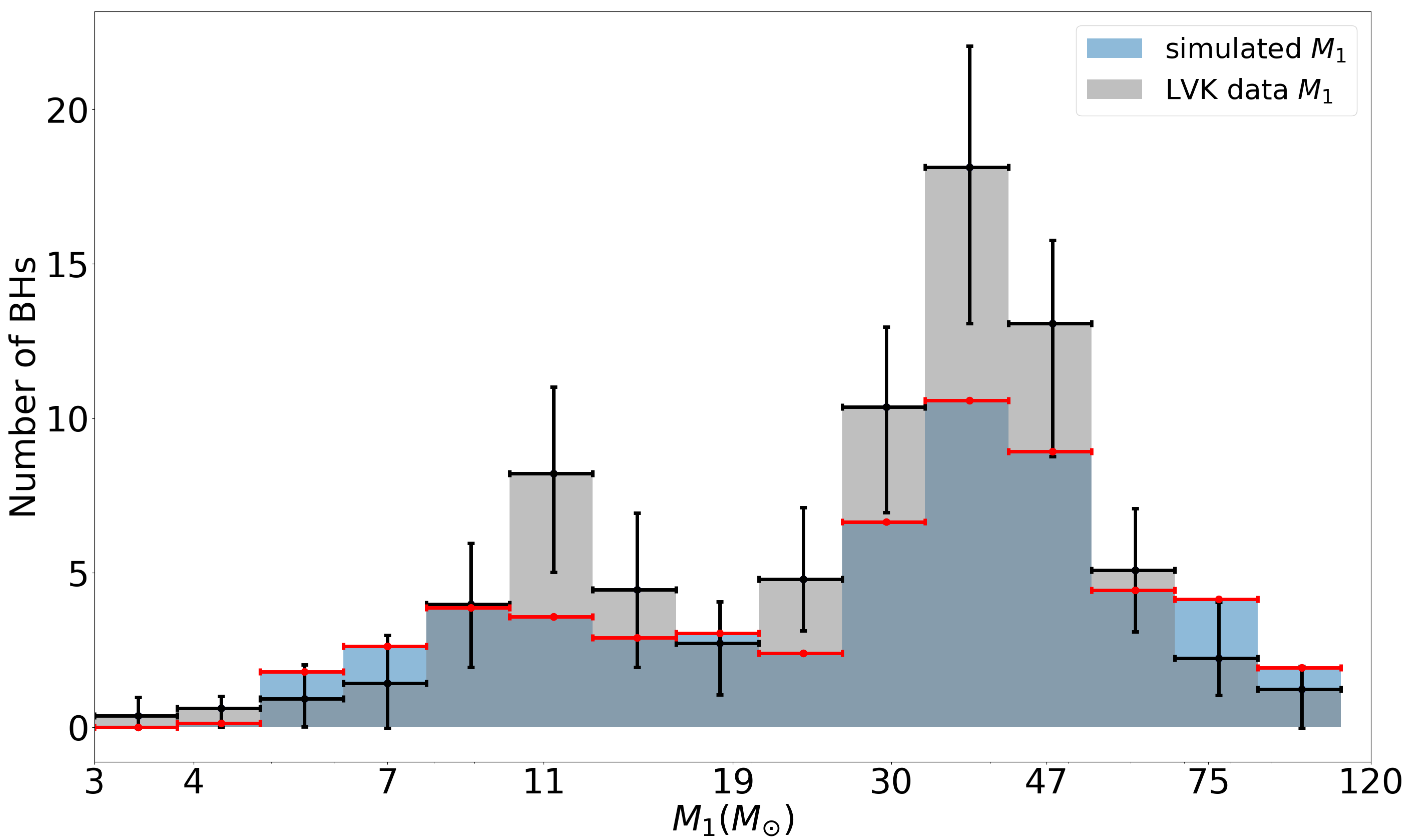}
    \caption{Normalized histogram of the distribution of $M_1$ of binary black holes with an $SNR>8$ that follow a PP distribution (blue columns with red normalizations). We plot here the case of $\alpha = 2.11$, $\beta=0$ and  $ \mu = 40M_\odot$. The corresponding standard deviation for a peak at $40M_\odot$ is $\sigma=6.38M_\odot$. The simulated BBHs are superposed to the histogram of the $M_1$ distribution from the LVK ones with Poisson error bars (grey columns).}
    \label{fig:opt_M_1_PP}
\end{figure*}

\begin{figure*}[!th]
    \centering
    \includegraphics[width=0.8\textwidth]{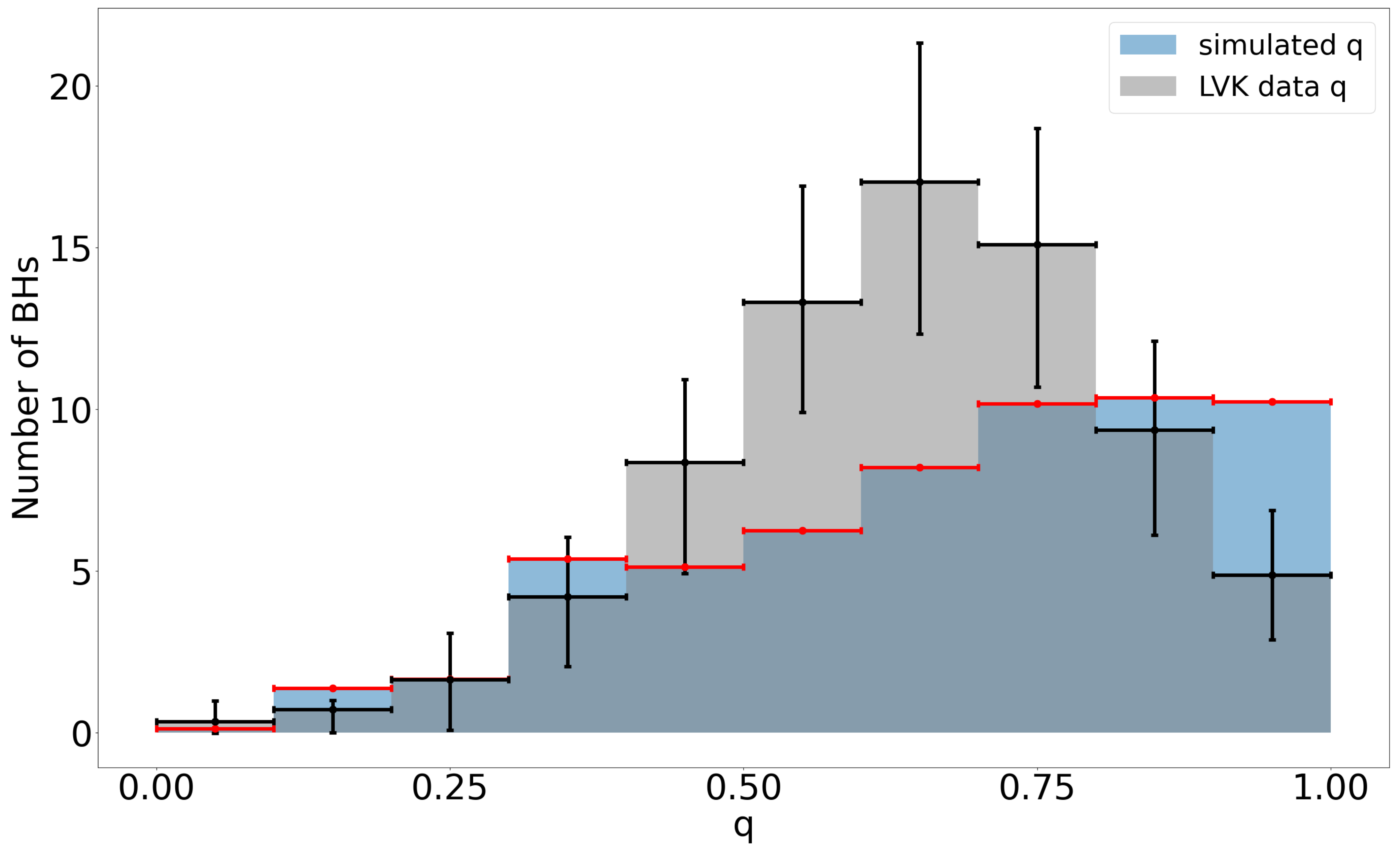}
    \caption{Normalized histogram of the distribution of $q$ of binary black holes with an $SNR>8$ that follow a power-law distribution (blue columns with red normalizations). We show results for  $\alpha = 2.11$, $\beta=0$ and  $\mu = 40M_\odot$. We also show the histogram of the $q$ distribution from the LVK BBHs with Poisson error bars (grey columns).}
    \label{fig:opt_q_PP}
\end{figure*}

Figs.~\ref{fig:opt_M_1_PP} and \ref{fig:opt_q_PP}, show the histograms of the simulated BHs from a power-law + Peak model, with the best-fit parameters compared to the histogram from the LVK data. 
The $M_1$ peak at around $40M_\odot$ seems to be insufficient in its amplitude when only looking at Fig.~\ref{fig:opt_M_1_PP}. However,
we remind the reader that when adding this peak component, 
we use of same normalization factor for both $M_1$ and $q$ (as it is the same binaries). 
The Peak component adds binaries almost entirely in the last $q\simeq 1$ bin.
Any attempt to make the Gaussian peak higher to fit the $M_1$ data better, also raises the number of counts in the last bin for the $q$-histogram. 

\begin{figure*}
\includegraphics[width=3.5in,angle=0]{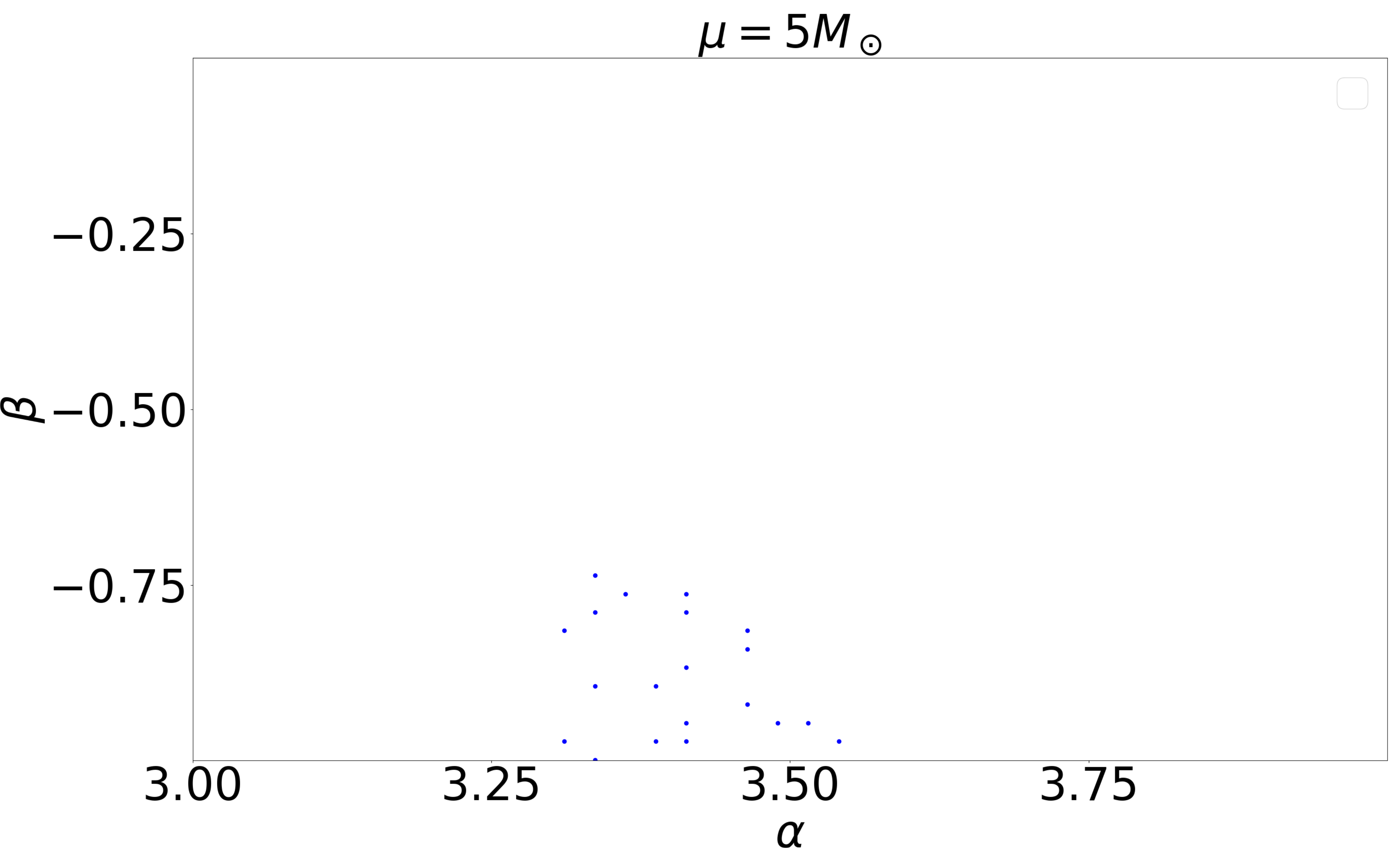} 
\includegraphics[width=3.5in,angle=0]{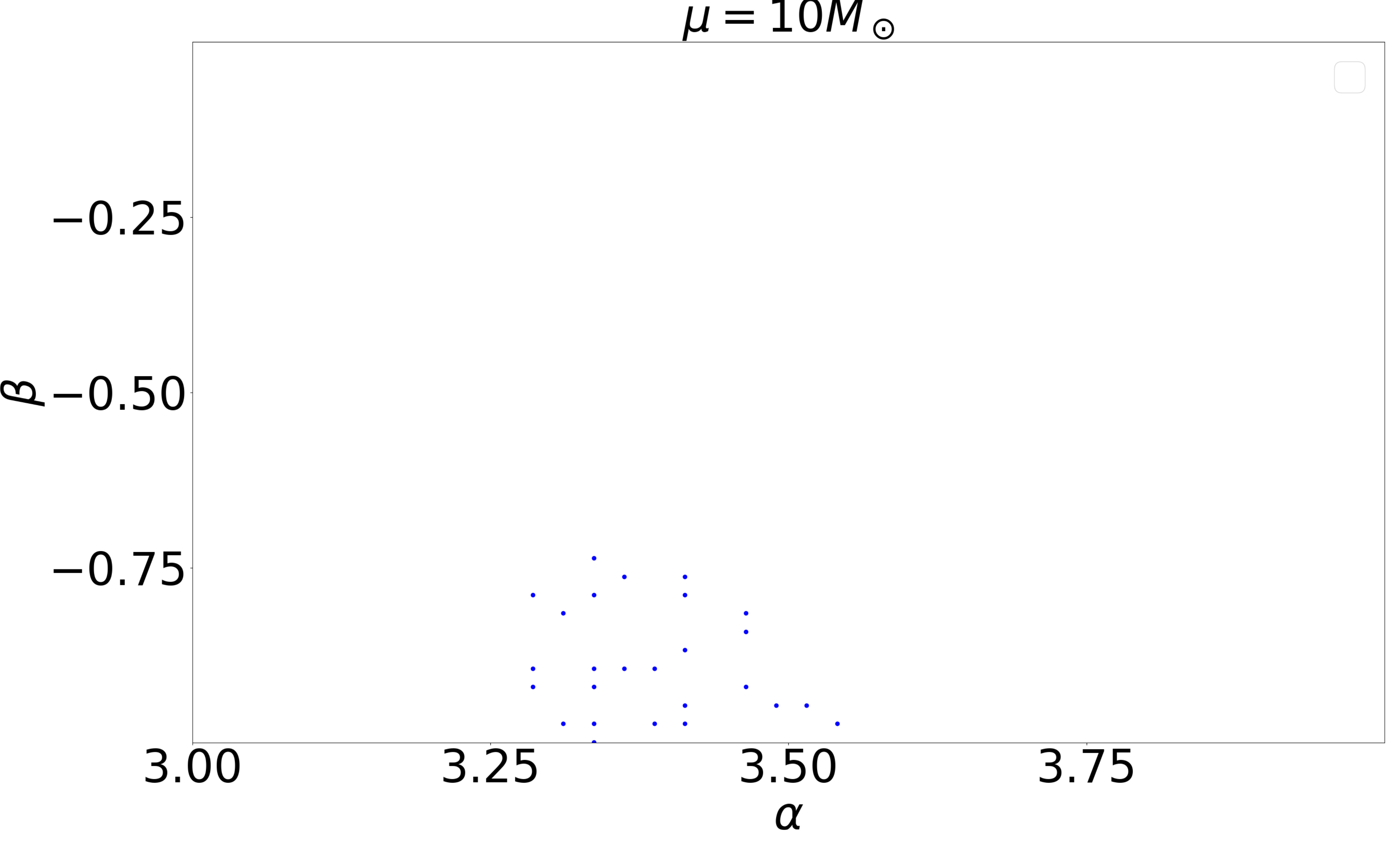}\\
\includegraphics[width=3.5in,angle=0]{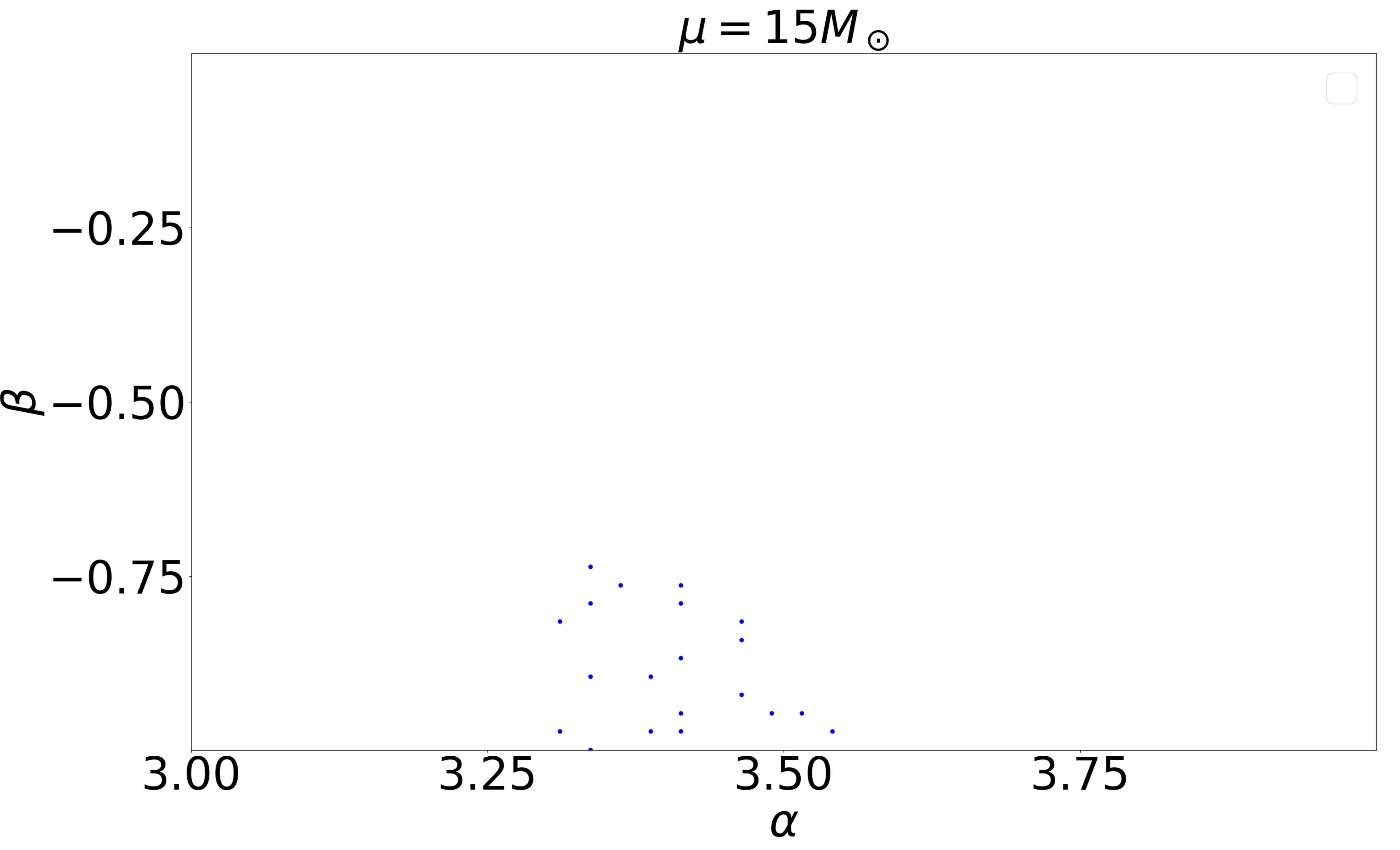} 
\includegraphics[width=3.5in,angle=0]{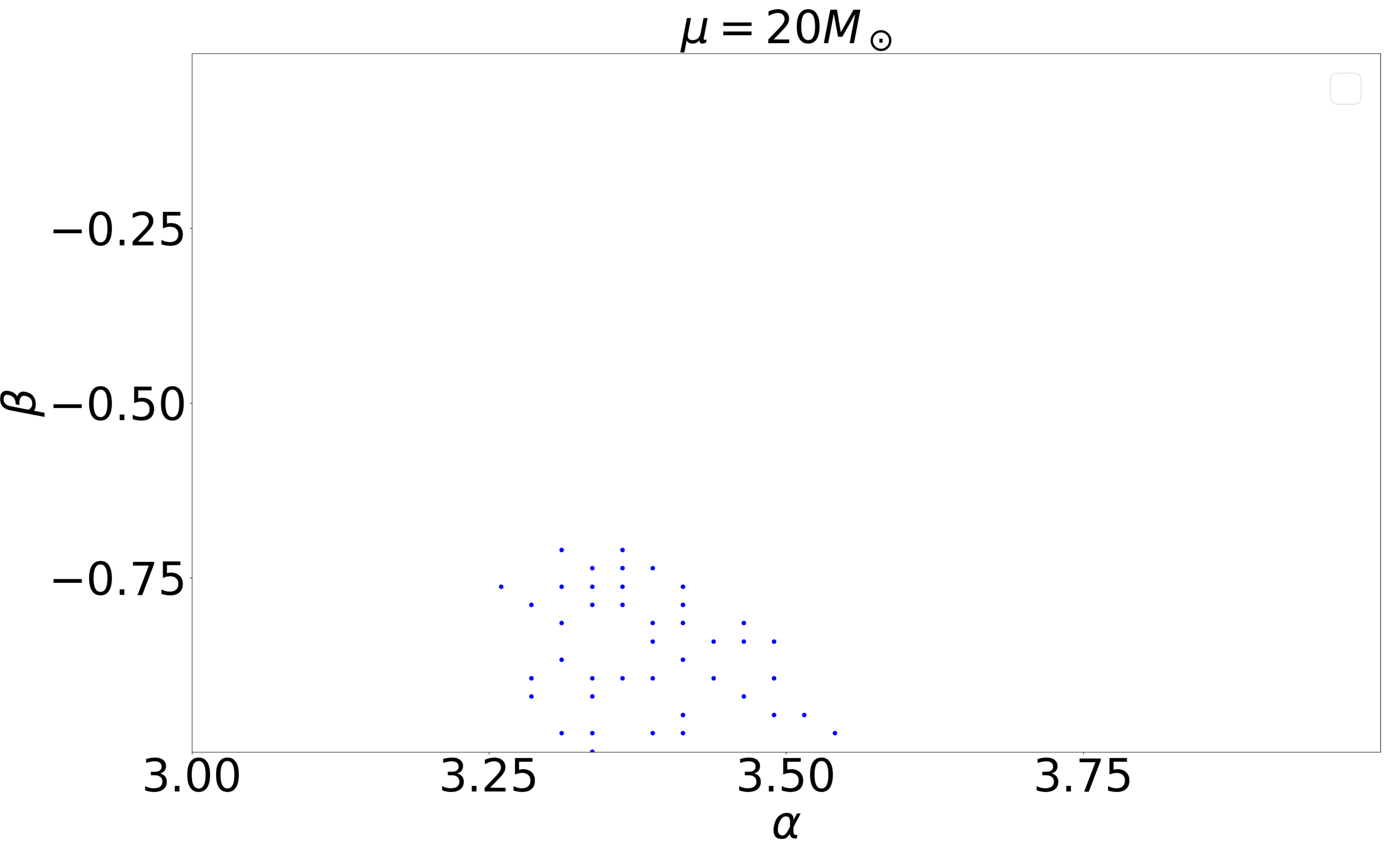}\\
\includegraphics[width=3.5in,angle=0]{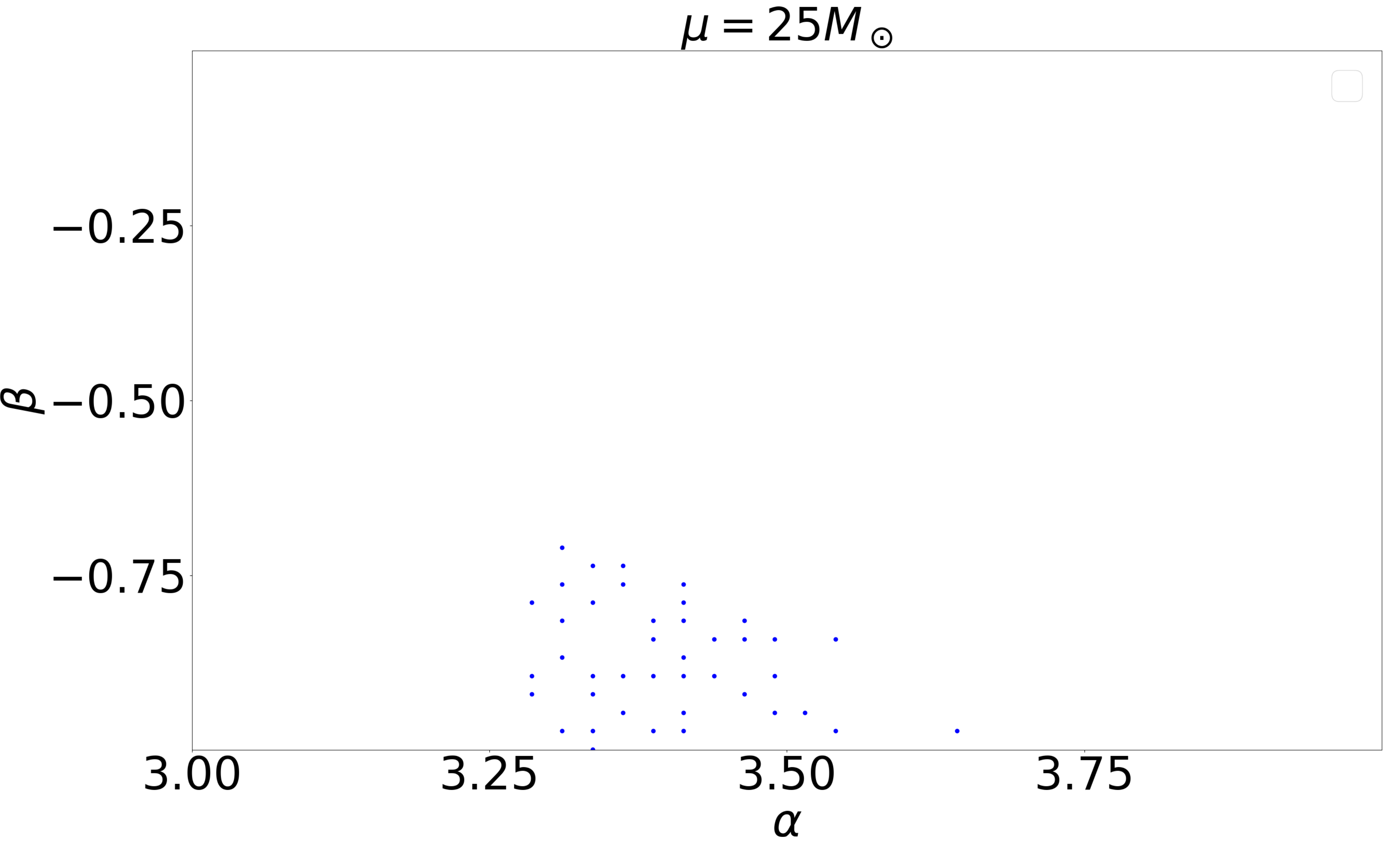} 
\includegraphics[width=3.5in,angle=0]{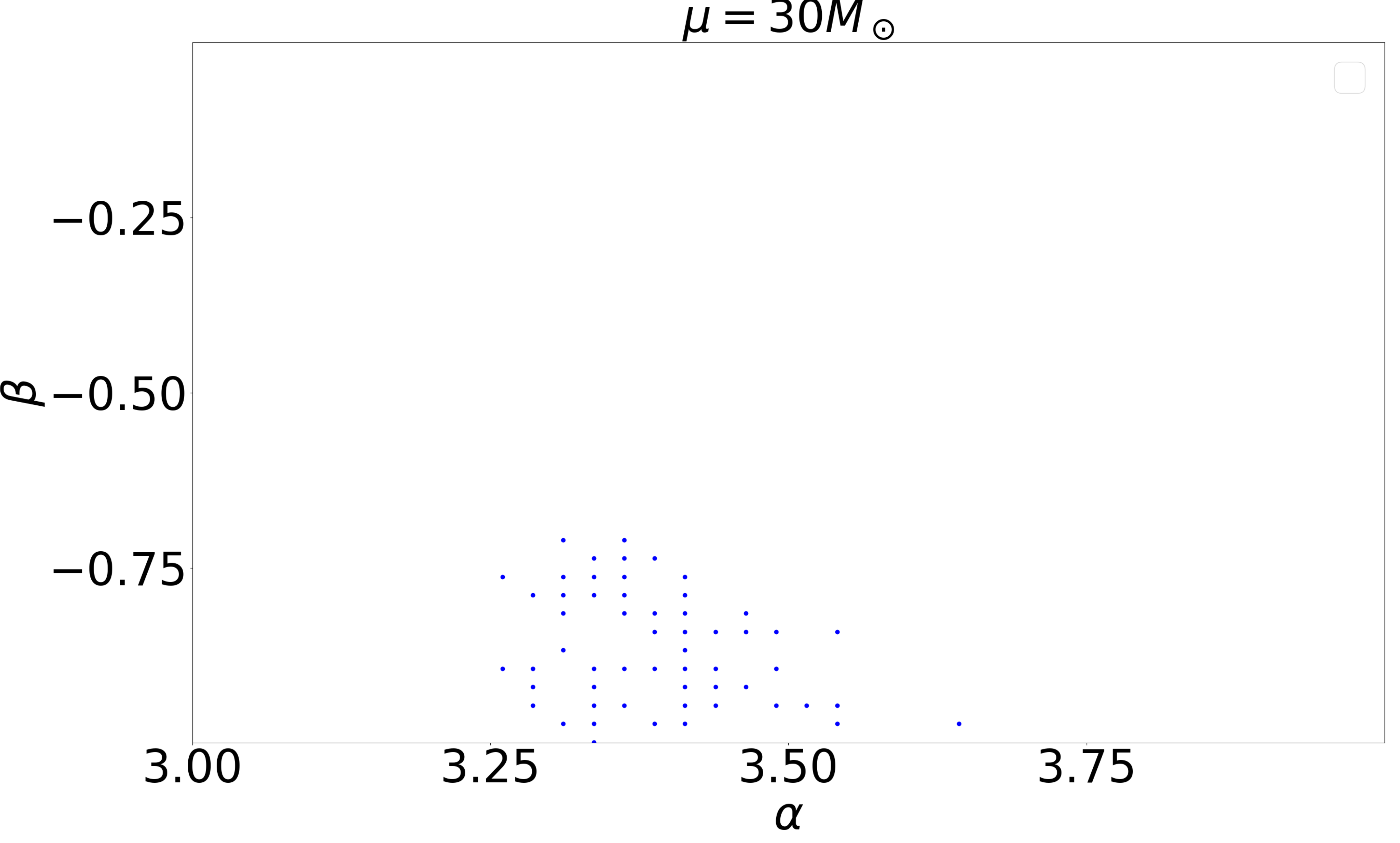}
\caption{The $\alpha$-$\beta$ parameter space for the power-law model probed using the BBHs from the LVK observations for different choices of $\mu$. The blue dots are inside the $90\%$ credible interval around the point giving the minimum $\chi^2$ value.}\label{fig:PP space param good chi 1}
\end{figure*}

\begin{figure*}
\includegraphics[width=3.5in,angle=0]{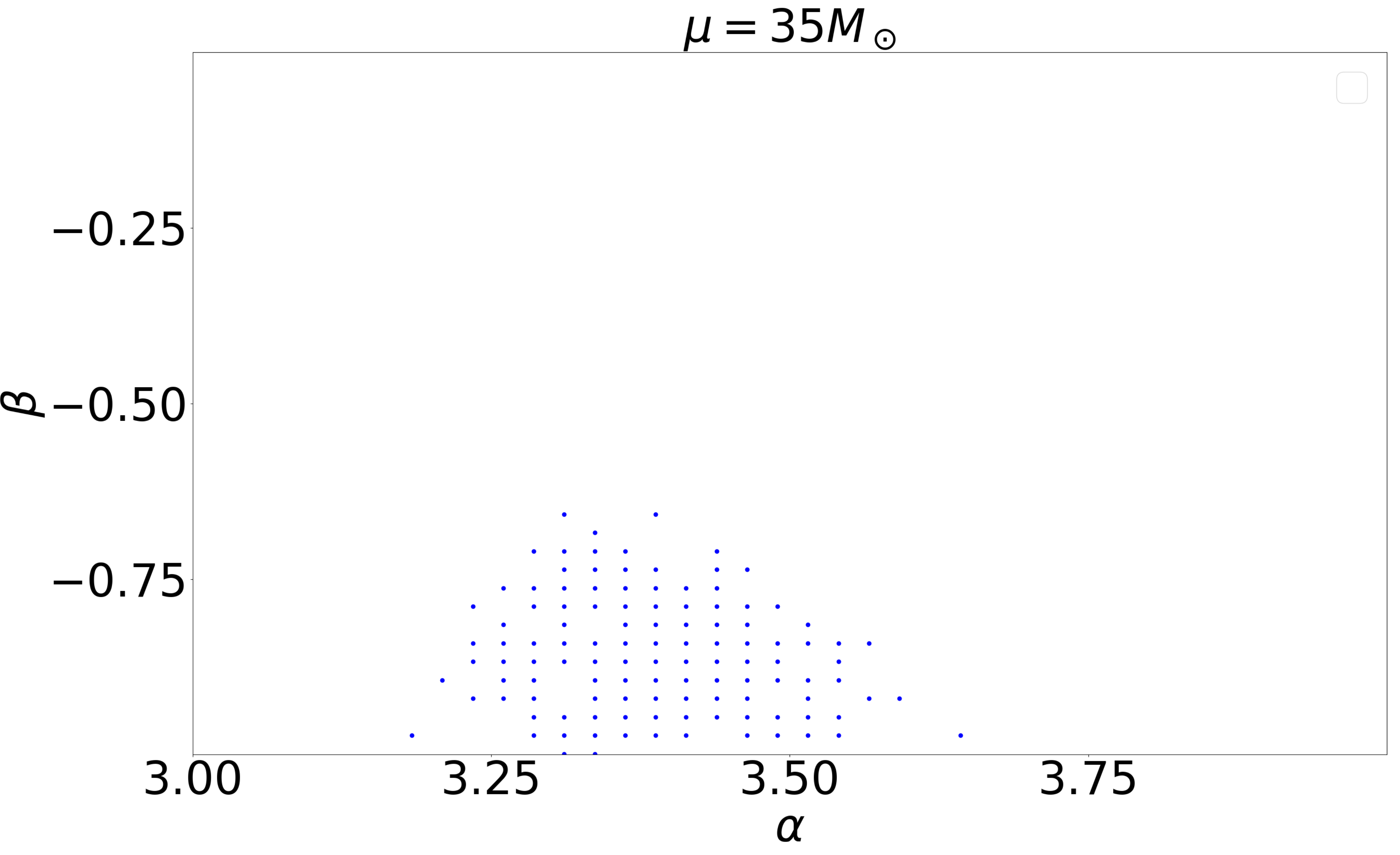}
\includegraphics[width=3.5in,angle=0]{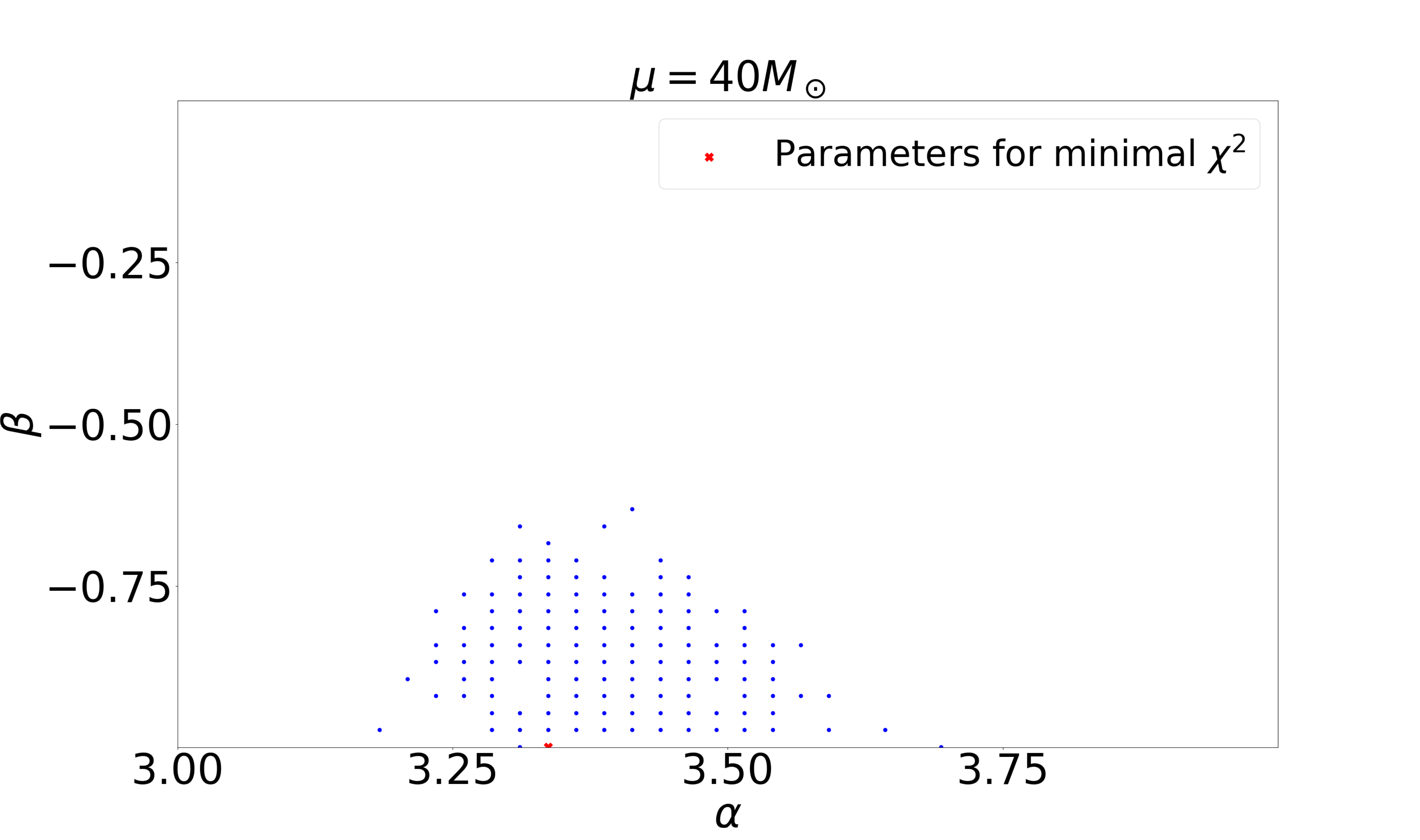}\\
\includegraphics[width=3.5in,angle=0]{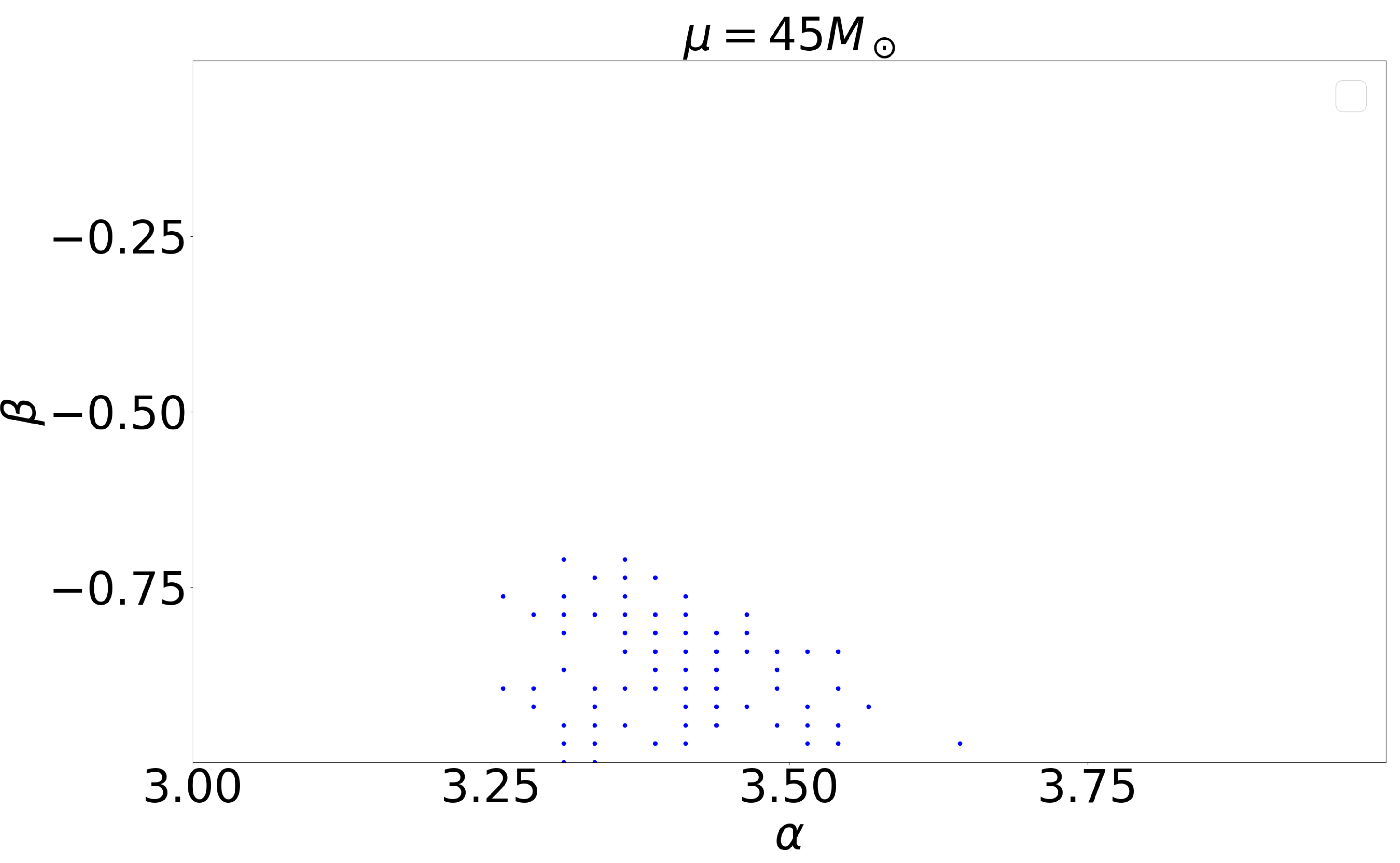}
\includegraphics[width=3.5in,angle=0]{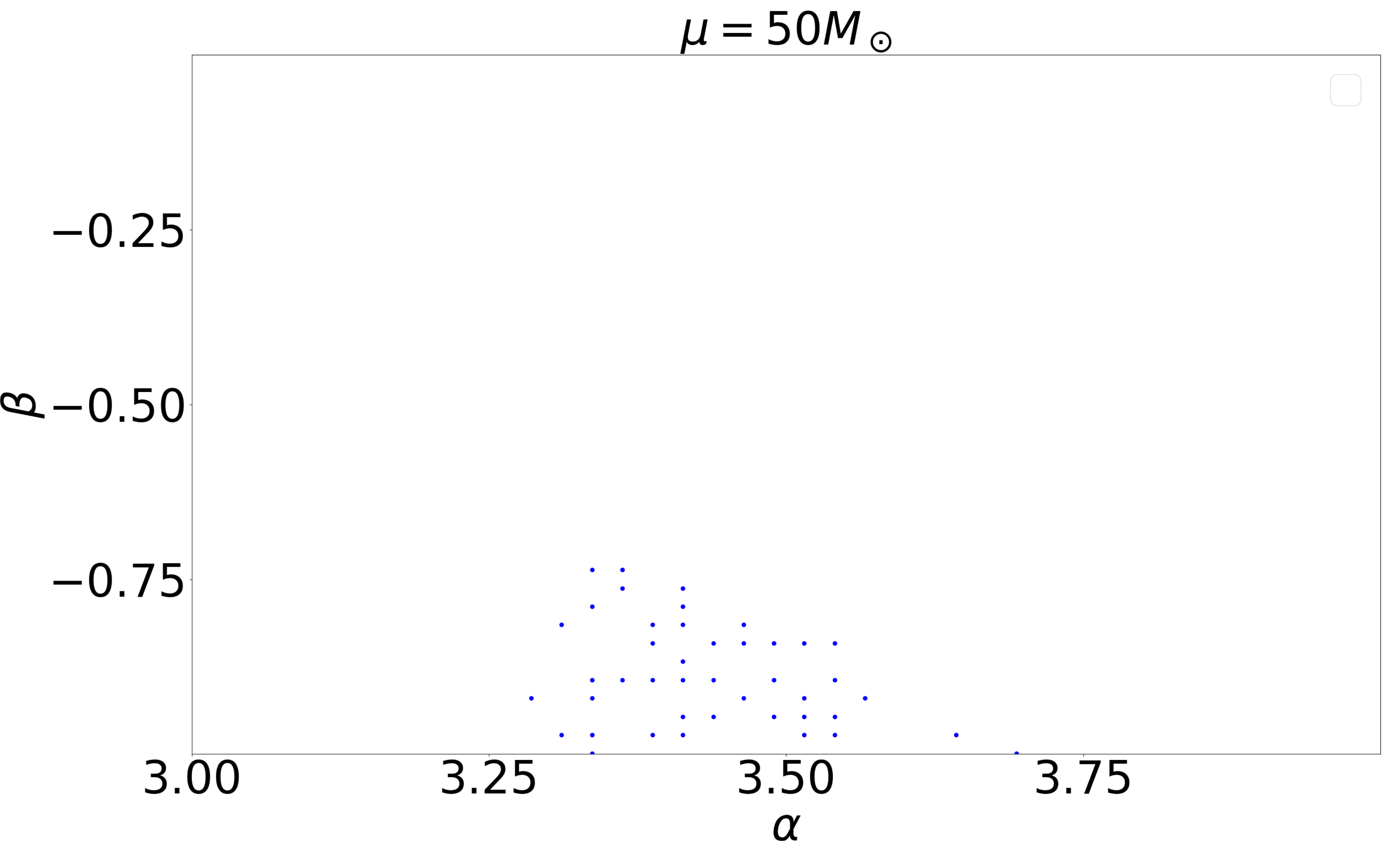}\\
\includegraphics[width=3.5in,angle=0]{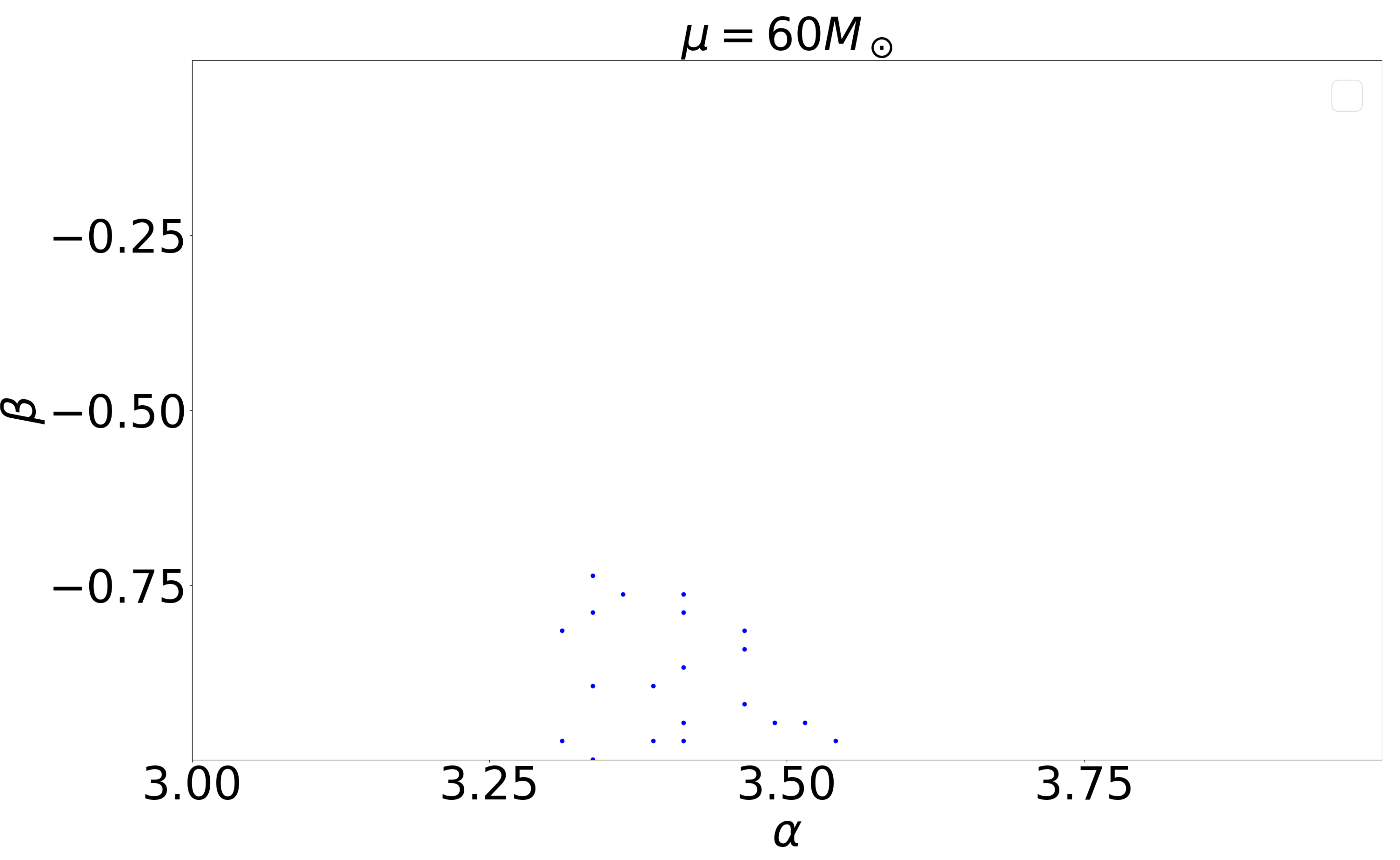} \includegraphics[width=3.5in,angle=0]{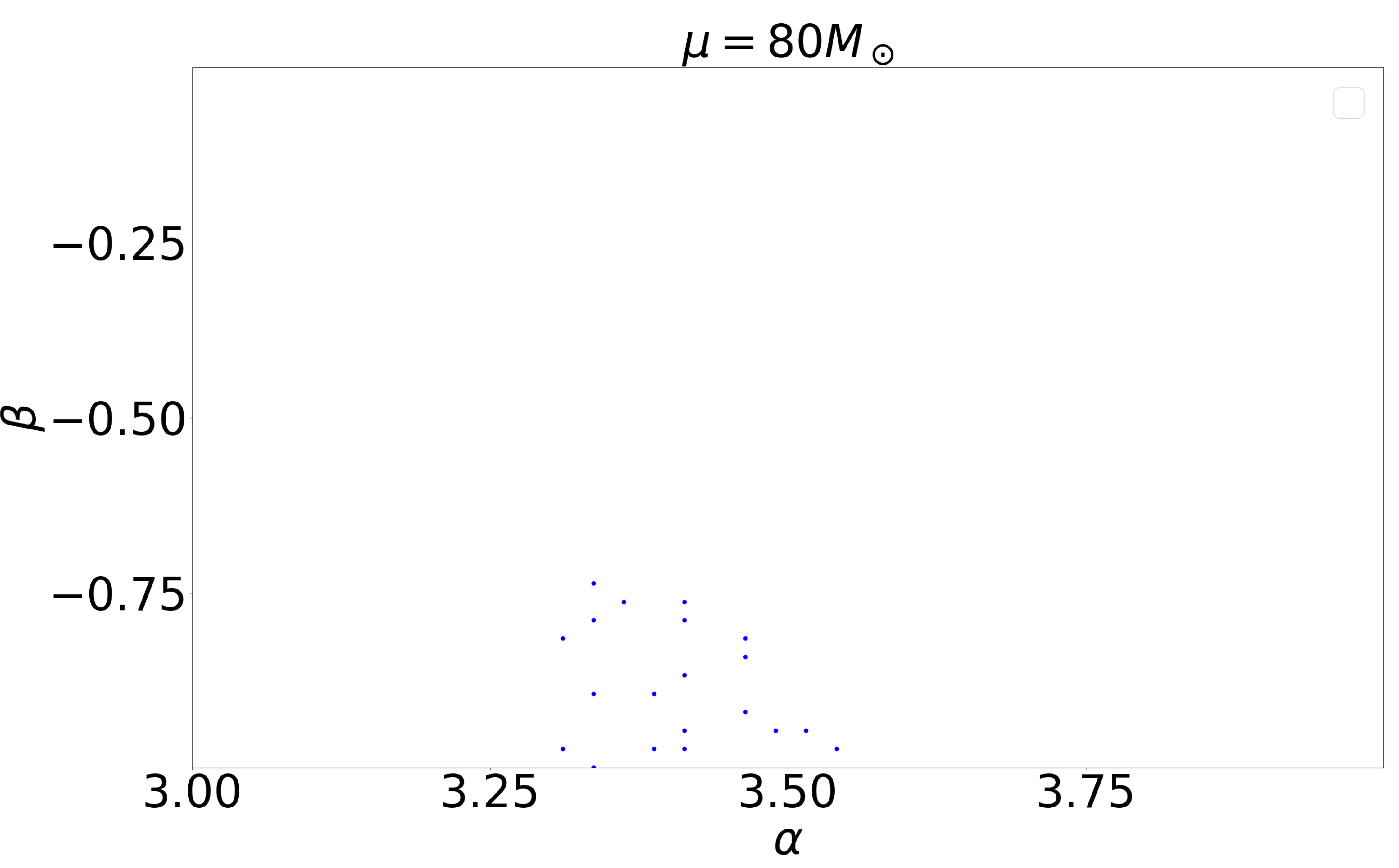}
\caption{As in Fig.~\ref{fig:PP space param good chi 1}, the $\alpha$-$\beta$ parameter space for the power-law model probed using the BBHs from the LVK observations for different choices of $\mu$. The blue dots are inside the $90\%$ credible interval around the point giving the minimum $\chi^2$ value. (red ``x'').}
\label{fig:PP space param good chi 2}
\end{figure*}

\begin{table}[th!]
\centering
\begin{tabular}{|c | c | c | c | c | c|} 
 \hline
 $\alpha$ & $\beta$ & $\mu$ & $\chi^2_{M_1}$ &  $\chi^2_q$ &  $\chi^2_{\textrm{d.o.f.}}$ \\ [0.5ex] 
 \hline
 3.34 & -1 & 40$M_\odot$ & 16.74 & 16.05 & 1.42\\ 
 \hline
  13.23& 0 & 40$M_\odot$ & 22.0 & 27.1 & 2.13\\
 \hline
  3.16& 1 & 40$M_\odot$ & 25.6 & 42.1 & 2.94\\ 
 \hline
   2.3& -0.45 & 35$M_\odot$ & 35.3 & 30.0 & 2.83\\ [1ex] 
 \hline
\end{tabular}
\caption{Results of the $\chi^2$ test after comparing simulated populations from the PP model to the 
LVK data. The number of degrees of freedom for this test is 23. The first line shows the best fit parameters and $\chi^2$ results, the second line results for $\beta=0$, the third line results for $\beta=1$and the last line the best fit results for $\alpha = 2.3$, which is the power-law exponent for the initial mass function of stars with masses > $3M_\odot$   Ref.~\cite{Kroupa_2001}.}
\label{table:chi_sqr_PP_results}
\end{table}
In Figs.~\ref{fig:PP space param good chi 1} and ~\ref{fig:PP space param good chi 2}, we show the area in parameter space around the best-fit values that is within $90\%$ confidence interval for the PP model. 
The $90\%$ confidence interval range for the PP model overlaps with the $90\%$ confidence interval of the power-law model. 
Some values of $\alpha$ and $\beta$ give a $\chi^2$ inside the $90\%$ confidence interval for the PP model, regardless of the normalization of the  Gaussian peak (even if very small). In these cases the PP fit ends up being effectively a power-law fit.\\

We did not probe values of $\beta < -1$ due to the unphysical nature of such a choice. Overall, we notice that the values of $\beta$ get smaller with an increase in the exponent of the redshift distribution $\kappa$. The redshift distribution exponent is quoted with a high uncertainty in Ref.~\cite{PhysRevX.13.011048}. Picking another value for $\kappa$ would result in a different $90\%$ confidence interval range for the PP and the PL models, but would still favor the PP model to the PL model in fitting the LVK data. We have tested $\kappa = 2.7$ as suggested by \cite{Madau:2014bja} and find that our results on $\alpha$ and $\beta$ do not change by more than $5\%$. We have also tested values of $\kappa$ as low as $\kappa=1$ and still find a preference for the PP over the PL model.


\section{Discussion and Conclusion} \label{sec:conc}
In this paper, we present a way of analyzing the LIGO-Virgo-KAGRA observations in terms of the mass and mass ratio spectra of BH binaries that it has detected. 
Taking each of the detected merger events of BBHs with a SNR$>8$, we construct $M_{1}$ and $q$ distributions as we have described in Section~\ref{sec:method}.
We then study how these distributions can be explained by different populations of black holes. 

We simulate a population of stellar-origin BHs. These BHs if they are dominated by first-generation BHs will have a mass distribution that \textit{only approximately} will resemble the mass-distribution of stars at their zero age (we describe that in Section~\ref{sec:Simulated_BBHs}).
We test a wide range of assumptions on those BHs' mass spectra. In each case we simulate a population of BBHs following specific $M_{1}$-, $q$- and $z$-distributions and evaluate how many of these binaries would be detected by the LVK collaboration, using the reported LIGO noise curves for the Livingston and Hanford observatories. We also allow for a second component of BBHs, where their mass spectrum would be approximated by a Gaussian distribution. Such a population could stem either from PBHs with a narrow mass-spectrum centred at the stellar mass range or from massive BHs that themselves are the product of earlier generation of BBH mergers. 

We find that the combination of both populations can much better explain the LVK observations (see our results in Section~\ref{sec:results}).
The best-fit choice to the $M_{1}$- and $q$-distributions that we study require for a stellar-origin BH population with a very similar spectrum to that expected for massive stars at their zero age, in combination with a Gaussian distribution centered at $40 M_{\odot}$.
Our result for the need of both components to explain the observations is in agreement with the claim made by Ref.~\cite{PhysRevX.13.011048}, but derived independently as we do not use information that could allow us to estimate rates of occurrence of individual merger events. We find somewhat different best fit values for the stellar-origin BH population's distribution properties. 
Our results agree fairly well with those of Ref.~\cite{PhysRevX.13.011048}, for the value of $\alpha$. However, we find a clear preference for a negative value of $\beta$, unlike~\cite{PhysRevX.13.011048}. Negative values of $\beta$, predict more binaries of unequal mass ratio. 
We consider our differences to be due to the inherently different way of representing the observations. With more detected events, as the statistical noise will decrease, we will be able to further understand the relevant systematic uncertainties of these two alternative methods of representing the LVK observations. 

Our method allows for a rapid analysis of the LVK observations as more results are soon to come from the O4 and future runs. In particular, it can be used to place limits on the merger rate of PBH binaries. We are currently working on the derivation of those limits using the Rates calculated in Ref.\cite{Aljaf:2024fru}.

We note that the models used in this paper are relatively simple, as the main aim of this work is to use the limited number of LVK BBHs as a showcase of how to address those questions (see also Ref.~\cite{Flitter:2020bky}). We leave for future work how to connect such an analysis to simulations of dense stellar environments, that would give us a prediction for the mass-properties of BHs coming from multiple/hierarchical mergers in them. This is of particular importance in relation to fitting the derived $q$-distribution. We noticed that neither a simple power-law BH mass-distribution nor a Gaussian component can optimally explain it. 

As a last note, we point out that for future more in-depth studies, LVK-detectors noise curves that account for the change in their sensitivity during a given observing run would be necessary. In this study we had to rely only on one noise curve per LIGO detector, derived during the early O3 run, considering it valid for the entire run. This hypothesis is a rough approximation, that can be easily improved on for future runs, if similar noise curves as those given in \cite{LIGOnoise}, could be provided for shorter time intervals during the observation period. 

\begin{acknowledgments}
IC and MEB are supported by the National Science Foundation, under grant PHY-2207912.
We would like to thank J. Hodaj and J. Decoste for useful discussions.

This research has made use of data or software obtained from the Gravitational Wave Open Science Center (gwosc.org), a service of the LIGO Scientific Collaboration, the Virgo Collaboration, and KAGRA. This material is based upon work supported by NSF's LIGO Laboratory which is a major facility fully funded by the National Science Foundation, as well as the Science and Technology Facilities Council (STFC) of the United Kingdom, the Max-Planck-Society (MPS), and the State of Niedersachsen/Germany for support of the construction of Advanced LIGO and construction and operation of the GEO600 detector. Additional support for Advanced LIGO was provided by the Australian Research Council. Virgo is funded, through the European Gravitational Observatory (EGO), by the French Centre National de Recherche Scientifique (CNRS), the Italian Istituto Nazionale di Fisica Nucleare (INFN) and the Dutch Nikhef, with contributions by institutions from Belgium, Germany, Greece, Hungary, Ireland, Japan, Monaco, Poland, Portugal, Spain. KAGRA is supported by Ministry of Education, Culture, Sports, Science and Technology (MEXT), Japan Society for the Promotion of Science (JSPS) in Japan; National Research Foundation (NRF) and Ministry of Science and ICT (MSIT) in Korea; Academia Sinica (AS) and National Science and Technology Council (NSTC) in Taiwan.

As part of this work we make publicly available
our primary mass and mass ratio data to make figures 2
and 3, in \url{https://zenodo.org/records/14675445}.
\end{acknowledgments}

\appendix






\bibliography{main.bib}


\end{document}